\documentclass[conference,compsoc]{IEEEtran}
\usepackage{hyperref}
\usepackage[nocompress]{cite}

\usepackage{booktabs}

\usepackage{enumitem}
\usepackage{booktabs}
\usepackage{multirow} 
\usepackage{graphicx} 
\usepackage{adjustbox}
\usepackage[table]{xcolor}

\usepackage{amsmath} 
\usepackage{amsfonts}

\usepackage{subcaption}
\usepackage{amssymb}

\usepackage{algorithm}
\usepackage{algorithmic}

\ifCLASSINFOpdf
\else
\fi
\begin{document}
%
\title{VLALeaks: Membership Inference Attacks against Vision-Language-Action Models}



%
\author{\IEEEauthorblockN{Xukun Luan\IEEEauthorrefmark{1},
Jinyan Liu\IEEEauthorrefmark{1},
Xuesong Li\IEEEauthorrefmark{1}, 
Yuanguo Bi\IEEEauthorrefmark{2},
Renjun Wu\IEEEauthorrefmark{1},
Zhongxiang Lei\IEEEauthorrefmark{1},
Di Wang\IEEEauthorrefmark{3}}
\IEEEauthorblockA{\IEEEauthorrefmark{1}School of Computer Science and Technology, Beijing Institute of Technology}
\IEEEauthorblockA{\IEEEauthorrefmark{2}School of Computer Science and Engineering, Northeastern University}
\IEEEauthorblockA{\IEEEauthorrefmark{3}PRADA Lab, King Abdullah University of Science and Technology}
Email: xukunluan@bit.edu.cn}



\maketitle

\begin{abstract}
Vision-Language-Action (VLA) models enable end-to-end robot control and have garnered widespread attention. However, the memorization of training data inherent to VLA, coupled with the high cost of robotic data acquisition, raises serious concerns regarding data privacy leakage and intellectual property infringement. Membership inference attacks (MIAs) aim to determine whether a given sample belongs to the training set. While representing a significant privacy threat, this attack remains underexplored in the context of VLA models. To bridge this gap, we propose VLALeaks, which is based on attention discrepancies in VLA models. We reveal, for the first time, the privacy vulnerabilities of VLA models. Specifically, it comprises a two-stage process: (1) membership feature extraction, and (2) attack model construction. Experimental results across multiple VLA benchmarks demonstrate that VLALeaks readily reveals membership information and achieves optimal attack AUC and TPR@1\%FPR, highlighting the privacy vulnerabilities in current VLA model deployments. Our work is the first systematic study of MIAs on VLA models, aiming to provide insights for secure and trustworthy VLA models\footnote{Our code is available at \url{https://github.com/Zili1000/VLALeaks}.}.
\end{abstract}


%
\IEEEpeerreviewmaketitle

\section{Introduction}
\label{Introduction}
Leveraging the success of pretrained foundation models, Large Language Models (LLMs), and Vision-Language Models (VLMs), Vision-Language-Action (VLAs) models have emerged to handle multimodal inputs—including visual observations and language instructions—and to produce robotic actions that accomplish embodied tasks~\cite{bu2025univla}. VLA models provide a crucial solution for robotic control and have been widely applied in industrial sorting, autonomous driving, and home assistants~\cite{ma2026survey}. By successfully bridging the gap between high-level instructions and the physical world, VLA models have attracted significant attention from both industry and academia. With the iterative development of VLA models such as RT-2~\cite{zitkovich2023rt}, $\pi_{0}$~\cite{black2024pi_0}, Gemini Robotics~\cite{team2025gemini}, $\Psi_0$~\cite{wei2026psi_0}, and OpenVLA~\cite{kim2024openvla}, robotic control has ushered in a new and more generalizable framework.

However, unlike traditional automated control (i.e., pre-programmed systems)~\cite{ze2025twist}, VLA models require large-scale, high-quality training sets to achieve satisfactory performance~\cite{wu2025robocoin,fei2025libero}. The high cost of collection and the complexity of annotation have led to an extreme scarcity of relevant VLA data. Furthermore, most available datasets are designed for specific or limited robot embodiments and configurations, offering little cross-platform or cross-embodiment generalizability. This characteristic introduces a new and underexplored privacy vulnerability. In particular, unlike the training data for LLMs, which can be readily scraped from the Internet, the data used to train VLA models often faces challenges such as expensive motion capture equipment, the need for professional motion guidance, and long data collection cycles~\cite{o2024open,lei2026kung}. These difficulties make VLA training data exceptionally valuable, thereby rendering its privacy protection even more critical. High-quality VLA data is remarkably expensive—comparable to gold coins~\cite{wu2025robocoin}. Membership inference attacks (MIAs)~\cite{shokri2017membership}, which aim to determine whether a given sample is contained in the training set, pose a significant privacy threat in this context~\cite{wang2025membership}.

Although traditional MIAs~\cite{hu2025membership,zaree2026attenmia,hayesexploring} have been extensively studied in the vision or language domains, they are either ineffective or exhibit low performance in the VLA context due to the following three limitations:
First, \textbf{Multimodal entanglement}: Vision, language, and action are aligned with one another, such that a difference in any single modality results in two distinct samples. This makes it difficult to analyze the boundary between member and non-member samples based on a single modality, thereby limiting direct threats to data privacy.
Second, \textbf{Constrained action output dimensionality}: VLA models typically restrict the output dimension (e.g., 7 × 256) according to the degrees of freedom of the robot, where the encoded 256-dimensional representation carries no actual semantic meaning. This renders methods that rely on output logits, confidence scores, or similarities ineffective for extracting membership information.
Third, \textbf{Minimal inter-sample variation}: Robotic manipulation tasks usually require hundreds of execution steps. During VLA data collection, the variation between consecutive frames is subtle, resulting in numerous highly similar samples. Consequently, MIAs on VLA models demand a more fine-grained extraction of membership information.

To address these challenges, we propose VLALeaks, the first MIA framework specifically designed for VLA models. VLALeaks introduces a novel two-stage inference strategy. In the first stage, the adversary leverages the VLA's self-attention mechanism to compute intra- and inter-modal membership features for information extraction. In the second stage, the adversary constructs an attack model using the features extracted in the previous stage to carry out the inference attack. VLALeaks captures membership information across the vision, language, and action modalities within VLA models, revealing a new privacy leakage channel.

Our contributions are as follows:

\begin{itemize}
    \item We provide the first formal characterization of membership privacy leakage risks on VLA models, a direction not explored in prior work.
    \item We propose VLALeaks, the first MIA framework specifically designed for VLA models. VLALeaks employs a two-stage attack strategy that extracts membership features hidden within the self-attention mechanisms of vision, language, and action modalities to perform inference attacks.
    \item We conduct extensive experiments on both simulated and real-world robotic platforms across multiple VLA models. Experimental results demonstrate that VLALeaks achieves SOTA attack AUC and TPR@1\%FPR.
\end{itemize}

In this paper, Section~\ref{Background} introduces the background and preliminaries. Section~\ref{Threat Model} describes the threat model. Section~\ref{Method} presents our method. Section~\ref{Experiments} presents the experiments. Section~\ref{Conclusion} concludes this work.

\section{Background and Preliminaries}
\label{Background}

\subsection{Vision-Language-Action}
Vision-Language-Action (VLA) models~\cite{ma2026survey} are typically built upon VLMs~\cite{liu2023visual}, enabling robots to comprehend human instructions and process visual inputs from cameras, thereby performing actions that adapt to the environment. VLA models achieve end-to-end control for robotic tasks. VLA models usually transform continuous action prediction into a classification problem by discretizing robot actions~\cite{kim2024openvla,zitkovich2023rt}. Specifically, the model first discretizes the continuous probability distribution into class labels $y = \arg \max \mathcal{F}(\boldsymbol{x})$, where $\mathcal{F}(\cdot)$ denotes the VLA model. Subsequently, an action de-tokenizer generates specific actions $\hat{x}_{act} = DT(y)$ based on these labels. By mapping action values to discrete class labels, the model converts continuous probability outputs into discrete signals. This simplification strategy helps accelerate convergence and reduce training time, and is therefore widely adopted in VLA-based robot models~\cite{black2024pi_0,wei2026psi_0}.

Acquiring VLA data typically involves expensive procedures such as collection and annotation. For instance, RoboCOIN~\cite{wu2025robocoin} utilizes teleoperation to collect over 180,000 demonstrations from 15 different robot platforms. Open X-Embodiment~\cite{o2024open} assembles a dataset from 22 distinct robots contributed by 21 institutions, covering 527 skills (160,266 tasks). Meanwhile, $\Psi_0$~\cite{wei2026psi_0} trains VLA models using large-scale first-person video data, supplemented by a small amount of human-robot interaction data. Collectively, these efforts underscore the paramount importance of high-quality data for VLA models, thereby highlighting the necessity of studying privacy and auditing for VLA data. Our proposed MIA effectively addresses this critical gap.

This study employs a robotic arm with 7 degree-of-freedoms (DoFs)~\cite{haddadin2022franka}. At each step, an action comprises 7 DoFs and has clear physical meanings in three-dimensional Cartesian space, represented as follows:
$Act = [\Delta P_x, \Delta P_y, \Delta P_z, \Delta R_x, \Delta R_y, \Delta R_z, \text{gripper}],$
where $\Delta P_{x,y,z}$ and $\Delta R_{x,y,z}$ denote the relative position changes and relative rotation changes along the $x$, $y$, $z$ axes, respectively. These changes are discretized into 256 uniform bins within the action interval $[y^i_{\text{min}}, y^i_{\text{max}}]$ for each degree of freedom. The gripper state is a binary variable indicating whether it is open or closed. This control mechanism introduces unique challenges for membership inference attacks: due to the extremely fine-grained bin division, the action differences between adjacent bins are very small (e.g., approximately $\pm 0.007$ units per bin), and the output space is constrained to $7 \times 256$. Therefore, more refined extraction methods are required for the membership characteristics of VLA model samples.

\subsection{Attention}

Transformer architecture~\cite{dosovitskiy2020image} is extensively employed in VLA models. The self-attention mechanism~\cite{vaswani2017attention} of the transformer models global token correlations using query, key, and value projections, serving as the fundamental building block for multimodal interaction modeling. Given an input feature sequence, the query, key, and value matrices are obtained through linear projections:
\begin{equation}
    Q = XW_Q,\quad K = XW_K,\quad V = XW_V,
\end{equation}
where $X$ denotes the input sequence feature, and $W_Q$, $W_K$, $W_V$ represent learnable projection parameters. The scaled dot-product is adopted to measure token-wise similarity, and the Softmax function normalizes the similarity into a valid probability distribution:
\begin{equation}
    A = \text{Softmax}\left(\frac{QK^\top}{\sqrt{d_k}}\right).
\end{equation}
Here, $d_k$ is the dimension of the key vector, and the scaling factor stabilizes the gradient magnitude during training. $A\in\mathbb{R}^{S\times S}$ denotes the standardized attention matrix, in which each element $A_{i,j}$ represents the attention weight from the $j$-th token to the $i$-th token. All weights satisfy the normalization constraint $\sum_j A_{i,j}=1$.

To enhance the capability of modeling diverse feature representations, multi-head attention splits the feature channels into multiple groups and performs parallel self-attention computations, followed by feature concatenation and fusion:
\begin{equation}
    \text{Head}_h = \text{Attention}(Q_h,K_h,V_h),\quad h=1,2,...,H,
\end{equation}
\begin{equation}
    \text{MultiHead} = \text{Concat}(\text{Head}_1,\text{Head}_2,...,\text{Head}_H)W_O,
\end{equation}
where $H$ denotes the total number of attention heads, and $W_O$ is the learnable output projection matrix for feature fusion. In this work, all attention-based analyses are conducted on the layer-wise averaged multi-head attention map:
\begin{equation}
    A^{(l)} = \frac{1}{H}\sum_{h=1}^{H} \text{Head}_h^{(l)}.
\end{equation}
$A^{(l)}$ denotes the aggregated global attention matrix of the $l$-th network layer, which serves as the basic input for the extraction of layer-wise characteristics and modeling of cross-layer dynamic evolution.

\subsection{Membership Inference Attacks}
Membership inference attacks (MIAs)~\cite{shokri2017membership,11544110,chen2025codepoisonMIA,du2026imitative,wang2025rigging} aim to determine whether a given sample belongs to the training set, which is a binary classification task. Existing works~\cite{wu2025membership,he2025towards} have extensively studied such attacks on various models. MIAs on LLMs target a single text modality and often rely on confidence scores or perplexity~\cite{zhang2025min} to extract membership features. Min-k~\cite{shi2024detecting} pioneers the investigation of MIAs against LLMs. Win-k~\cite{wink} researches the first MIAs against small language models. Deng \textit{et al.}~\cite{deng2026efficientmembershipinferenceattacks} propose ProjRes, the first projection residuals-based passive MIA tailored for federated large language models. MIAs on VLMs need to consider both vision and language modalities. For instance, MaxRényi~\cite{li2024membership} and KCMP~\cite{yin2026black} perform attacks on image or text modalities separately. Hu \textit{et al.}~\cite{hu2025membership} propose five types of image-text pair MIAs targeting VLMs during fine-tuning. VidLeaks~\cite{wang2026vidleaks} first investigates membership privacy leakage in the context of text-to-video models. Wang \textit{et al.}~\cite{wang2026inference} propose MIAs targeting graph generative diffusion models. Pang \textit{et al.}~\cite{pang2023black} propose the first scores-based MIA framework tailored for fine-tuned diffusion models. Hu \textit{et al.}~\cite{hu2026reasoning} explore privacy leakage of training data members in large reasoning models. To the best of our knowledge, MIAs on VLA models remain largely underexplored. This is particularly concerning given that the privacy leakage risks associated with VLA models are increasing by the day. We argue that the internal attention matrices of VLA models may provide more discriminative membership features across the three modalities—vision, language, and action.

\begin{figure}
    \centering
    \begin{subfigure}[b]{0.9\linewidth}
        \centering
        \includegraphics[width=\linewidth]{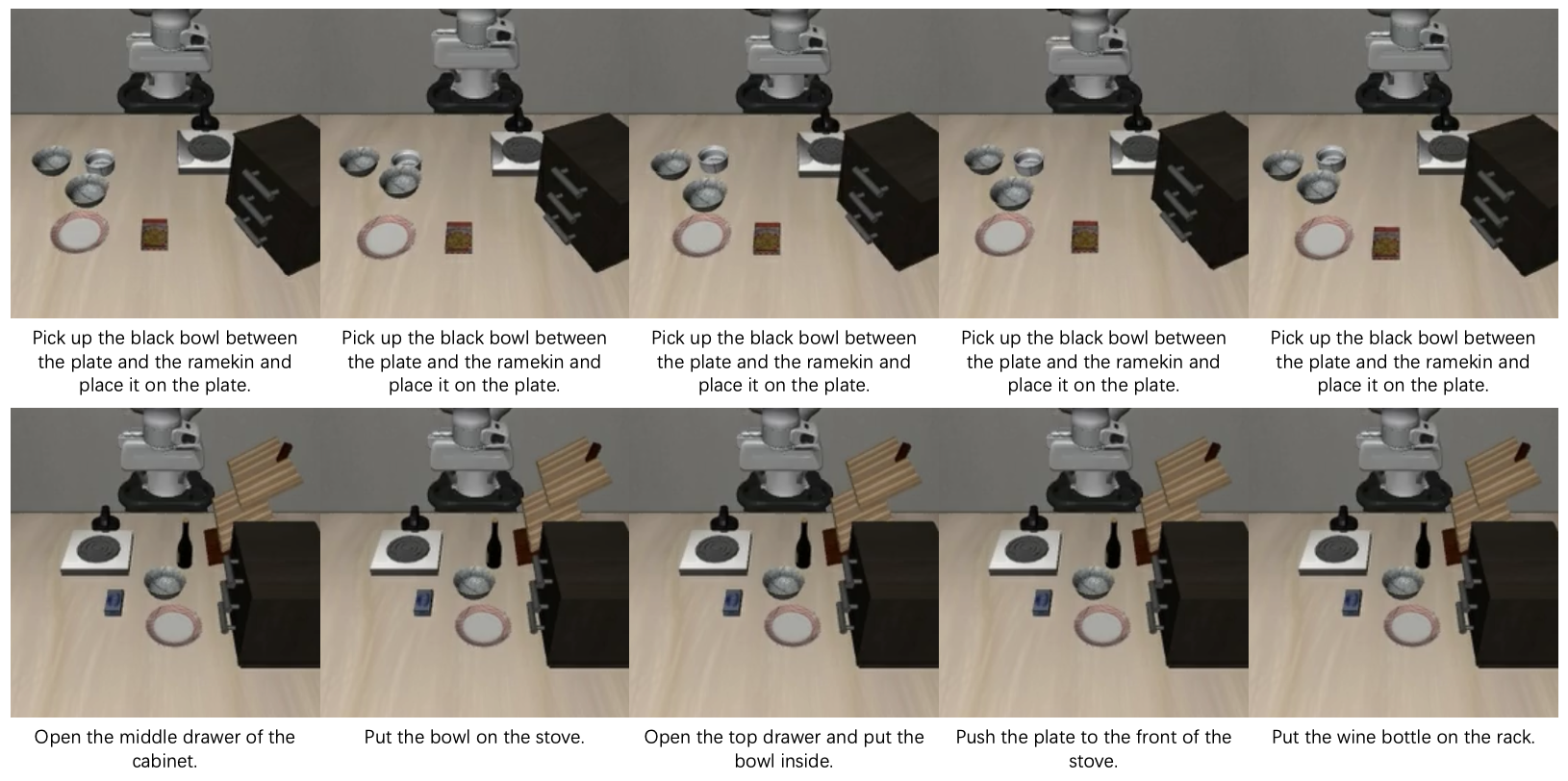}
        \caption{Subtle differences in images, but identical text; identical images, varying text.}
        \label{fig:VL}
    \end{subfigure}
    \hfill
    \begin{subfigure}[b]{0.9\linewidth}
        \centering
        \includegraphics[width=\linewidth]{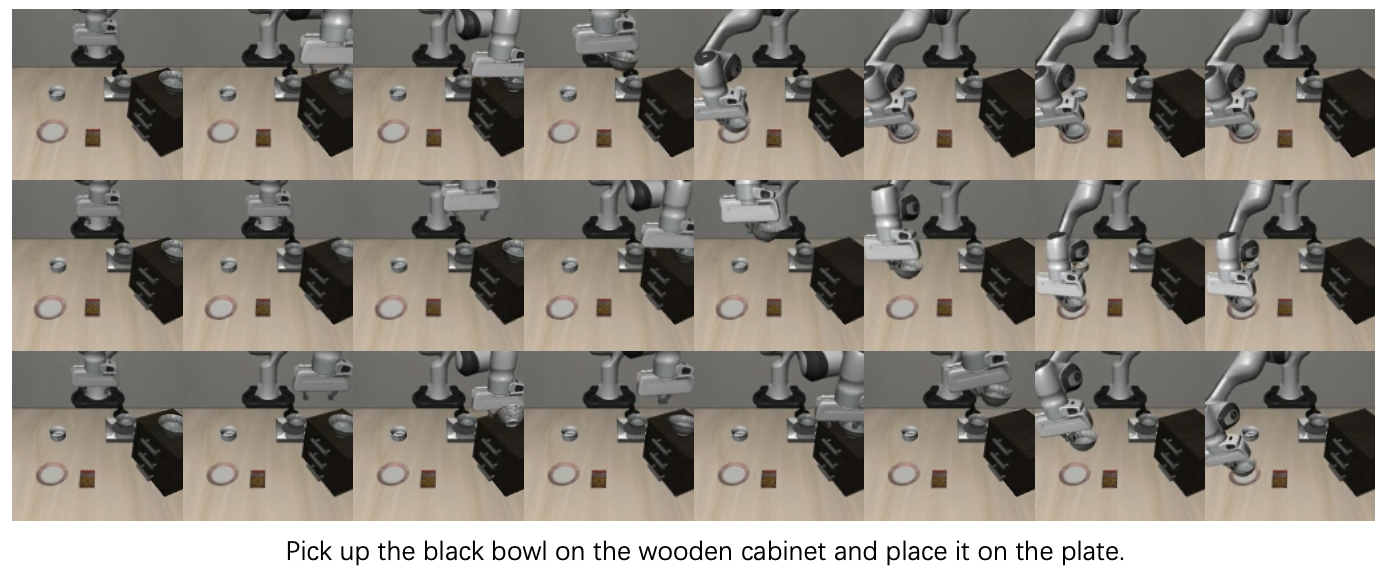}
        \caption{Given identical text and the same initial image, the actions taken when performing the task differ.}
        \label{fig:A}
    \end{subfigure}
    \caption{Characteristics of VLA data from VLA models pose unique challenges for MIAs.}
    \label{fig:data}
\end{figure}

As with the existing MIAs discussed above, analyzing the unique properties of VLA data is beneficial for guiding algorithm design. Figure~\ref{fig:data} summarizes three intrinsic unique properties of VLA data:
(1) identical text paired with visually distinct input scene images;
(2) unchanged observation images matched against different task texts;
(3) fully consistent initial image-text input yet divergent robot manipulation action trajectories.
These three irregular cross-modal mismatches break the strict one-to-one correspondence among vision, language, and action modalities that conventional MIAs rely on. Classic MIAs are originally designed for unimodal or well-aligned multimodal models, which assume fixed mapping between input and model outputs. When applied to VLA models, the unstable input-output correlation caused by the above three dataset characteristics severely damages the feature discrimination of traditional attacks and degrades their inference accuracy, naturally explaining the poor performance of existing MIA baselines (Section~\ref{attackauc}) in our quantitative experiments. In contrast, our proposed VLALeaks is tailored to mine modality-agnostic implicit privacy hidden amid such unaligned VLA data, thus achieving remarkable attack superiority over prior arts.

\subsection{Formulation of MIAs to VLA models}
Let $\mathcal{F}(\cdot)$ denotes a trained VLA model. The VLA model takes a triple-modal sample $\boldsymbol{x}=(x_{img},x_{txt},x_{act})$, where $x_{img}$ denotes visual observation, $x_{txt}$ denotes language instruction, and $x_{act}$ denotes action-related information. The model outputs a predictive action distribution $\mathcal{F}(\boldsymbol{x})$.
We define $\mathcal{D}_{\text{train}}$ as the private training dataset (member) of the target VLA model and $\mathcal{D}_{\text{test}}$ as the public non-training dataset (non-member), where $\mathcal{D}_{\text{train}} \cap \mathcal{D}_{\text{test}} = \emptyset$.

We define a binary attack function:
\begin{equation}
\mathbb{AF}(\boldsymbol{x})=
\begin{cases}
1, & \boldsymbol{x} \in \mathcal{D}_{\text{train}} \quad \text{(member)}\\
0, & \boldsymbol{x} \in \mathcal{D}_{\text{test}} \quad \text{(non-member)}.
\end{cases}
\end{equation}
In contrast to classical MIAs~\cite{wu2025membership} on unimodal models (e.g., image classifiers or text encoders), the adversary must specifically consider three heterogeneous modalities (vision, language, and action) simultaneously. The interaction among these modalities introduces unique challenges: the attack model must capture cross-modal correlations, handle potential modality-wise overfitting, and exploit leakage that may arise from any single modality or their joint representation. Consequently, effective MIAs against VLA models require not only traditional membership features but also modality-specific and fused features that reflect how the VLA model integrates multi-modal information during training and inference.

\section{Threat Model}
\label{Threat Model}
\subsection{Adversary's Goal}
\label{Adversary's Goal}

The adversary aims to compromise the membership privacy of VLA models. Given a target sample $\boldsymbol{x}$ to be examined, the adversary attempts to perform binary classification. The attack performance improves as the adversary more accurately classifies samples from the training set as 1 and samples not in the training set as 0.

\subsection{Adversary's Knowledge}

We assume the adversary has white-box access to the VLA model. Specifically, the adversary can access the model's architecture, internal parameters, and attention mechanisms. This is a realistic assumption, as many VLA models today are open-sourced (e.g., OpenVLA~\cite{kim2024openvla}, SpatialVLA~\cite{qu2025spatialvla}, $\pi_{0}$~\cite{black2024pi_0}), and developers often fine-tune these models for downstream tasks. White-box access is a common setting in the field of VLA model security~\cite{zhou2026badvla,wang2025exploring,li2026attentionbetrayserasingbackdoor}.

\subsection{Adversary's Capability}

The adversary can only operate during the inference phase of the model. Specifically, the adversary feeds the target sample forward through the VLA model and obtains the output attention weight matrix. Formally, for a target sample $\boldsymbol{x}$, the adversary can extract the attention weight matrix $A$, which is then used to extract membership features.

\section{Method}
\label{Method}
\begin{figure*}
    \centering
    \includegraphics[width=0.96\linewidth]{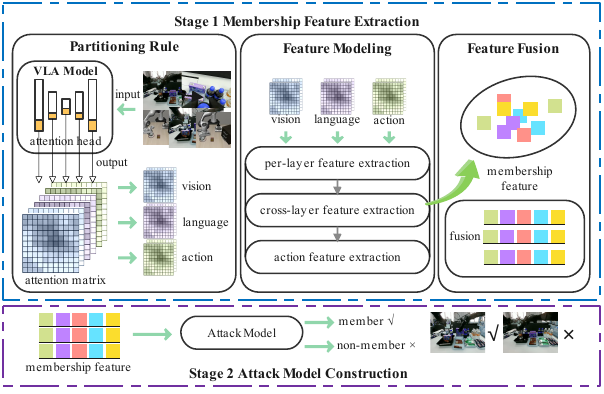}
    \caption{Overview of VLALeaks. VLALeaks comprises a two-stage inference pipeline. In Stage 1, Membership Feature Extraction, the method extracts attention matrices from the VLA model for a given input. These matrices are then processed via Partitioning Rule, Feature Modeling, and Feature Fusion to construct membership features. In Stage 2, Attack Model Construction, a binary classifier (model) is trained on these features to perform membership inference. We validate our approach on both virtual simulation and real-world data.}
    \label{fig:head}
\end{figure*}

We propose a two-stage MIA framework, VLALeaks, to compromise the membership information of VLA models in Figure~\ref{fig:head}. As illustrated in Section~\ref{Adversary's Goal}, VLALeaks aims to determine whether a given sample $\boldsymbol{x}$ belongs to the training set. As shown in Algorithm~\ref{alg:vla_leaks}, the two-stage inference process consists of: (1) membership feature extraction, where statistical membership features are computed from the model's attention matrices; and (2) attack model construction, where a binary classifier is trained using the extracted membership features.

\subsection{Stage 1: Membership Feature Extraction}

\begin{figure*}
    \centering
    \begin{subfigure}[b]{\linewidth}
        \centering
        \includegraphics[width=0.76\linewidth]{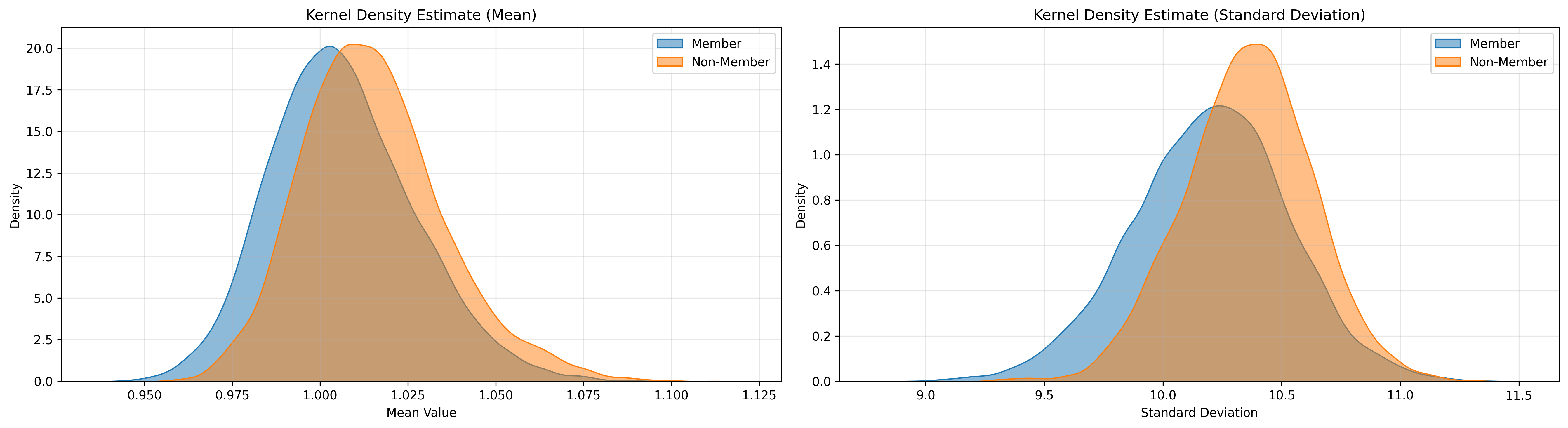}
        \caption{LIBERO-100}
        \label{fig:kde_100}
    \end{subfigure}
    \hfill
    \begin{subfigure}[b]{\linewidth}
        \centering
        \includegraphics[width=0.76\linewidth]{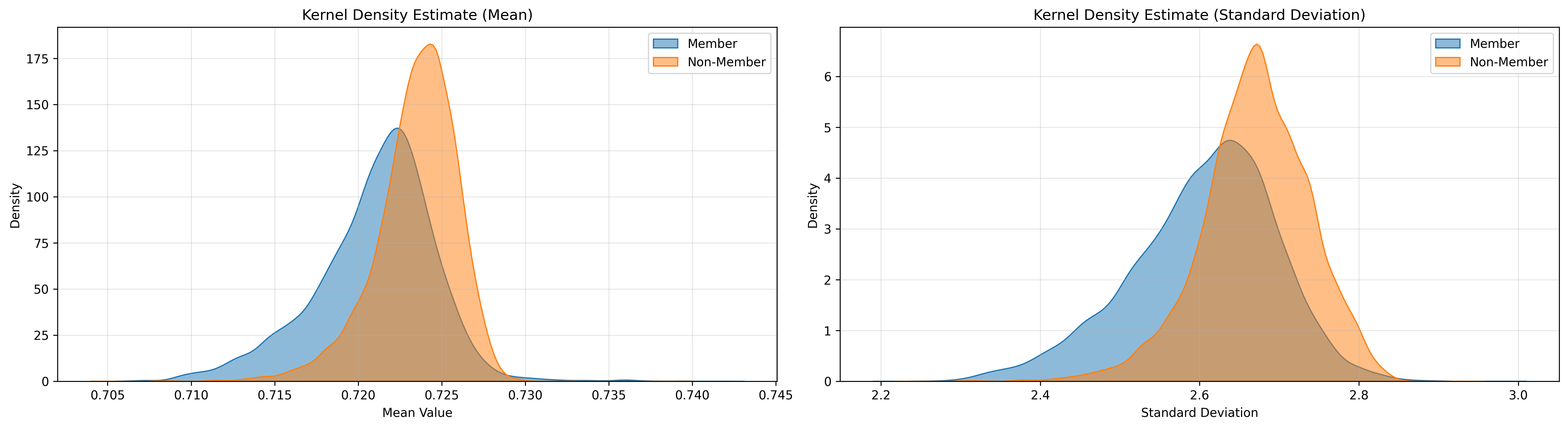}
        \caption{LIBERO-Goal}
        \label{fig:kde_goal}
    \end{subfigure}
    \caption{Kernel density estimates of attention features for members (blue) and non-members (orange). We evaluate the mean and standard deviation of OpenVLA~\cite{kim2024openvla} on LIBERO-100/Goal~\cite{liu2023libero}. The feature modeling process captures membership features of the VLA models, which are then used for inference in Stage 2.}
    \label{fig:kde}
\end{figure*}

The key intuition behind VLALeaks is that member and non-member samples exhibit differences in the attention matrices across vision, language, and action modalities. As shown in Figure~\ref{fig:kde}, we evaluate the mean and standard deviation of the attention matrices on OpenVLA~\cite{kim2024openvla}. We observe that member samples exhibit a more concentrated distribution compared to non-member samples. This discrepancy is the key to the success of VLALeaks. To align with the structural characteristics of VLA models, we design a membership feature extraction scheme.

Specifically, VLALeaks takes as input the multi-layer attention weight matrices of a VLA model and outputs membership features that integrate information from the vision, language, and action modalities.

\begin{enumerate}
    \item Partition the attention weight matrix into three modality-specific token interaction regions (vision, language, and action) based on a sequence semantic partitioning rule.
    \item Compute, for each layer, statistical distribution features of the attention, cross-layer evolution features, and intra-modal/inter-modal interaction features.
    \item Aggregate all attention features and output a multi-dimensional attention feature set, which serves as the membership feature for subsequent attack execution.
\end{enumerate}

\subsubsection{Partitioning Rule}

The attention matrix is divided into three distinct regions corresponding to the three modalities:

\begin{itemize}
    \item Vision modality region: The first $S_{img}$ tokens of the sequence, corresponding to sub-matrix $A_{img}^{(l)}$.
    \item Language modality region: The segment between the image tokens and the action tokens, corresponding to language sub-matrix $A_{txt}^{(l)}$.
    \item Action modality region: A fixed interval at the end of the sequence, corresponding to action sub-matrix $A_{act}^{(l)}$.
\end{itemize}

\subsubsection{Feature Modeling}

Feature modeling is the core component of VLALeaks, comprising: (1) per-layer single attention layer feature extraction, (2) cross-layer temporal evolution feature extraction, and (3) action-modality-specific feature extraction.

(1) Per-Layer Single Attention Layer Feature Extraction. We further decompose the vision and language modalities into four sub-interaction regions: intra-image interaction $A_{img}^{(l)}$, image-text cross-modal interaction $A_{img2txt}^{(l)}$, text-image cross-modal interaction $A_{txt2img}^{(l)}$, and intra-text interaction $A_{txt}^{(l)}$. VLALeaks iterates through the attention matrix of each layer of the model and independently computes basic representations for each of the four decomposed interaction sub-regions. The feature dimensions extracted at a single layer include:

\begin{itemize}
    \item Magnitude statistical features: Mean and standard deviation of attention weights in the region, characterizing interaction intensity and dispersion.
    \item Distribution entropy features: Attention entropy, representing the degree of attention diffusion — higher entropy indicates more dispersed interaction.
    \item Focusing degree features: Maximum attention weight ratio, measuring the model's ability to concentrate attention on a single point.
\end{itemize}

Statistical, distributional, and focusing features are calculated for single-layer attention sub-matrices. Let $\mathcal{A}$ denote any attention sub-matrix, and $N$ denote the total number of elements in the sub-matrix.

\noindent \textbf{Attention Mean (Interaction Intensity)}
\begin{equation}
\mu(\mathcal{A}) = \frac{1}{N}\sum_{i}\sum_{j}\mathcal{A}_{i,j}.
\end{equation}
\noindent \textbf{Attention Standard Deviation (Distribution Dispersion)}
\begin{equation}
\sigma(\mathcal{A}) = \sqrt{\frac{1}{N}\sum_{i}\sum_{j}\big(\mathcal{A}_{i,j}-\mu(\mathcal{A})\big)^2}.
\end{equation}
\noindent \textbf{Attention Entropy (Distribution Diffusion)}

Single-row attention entropy:
\begin{equation}
H(\textit{a}) = -\sum_{k} a_k \log(a_k+\varepsilon).
\end{equation}

Regional average attention entropy:
\begin{equation}
H_{avg}(\mathcal{A}) = \frac{1}{N}\sum_{i} H(\mathcal{A}_{i,:}).
\end{equation}
\noindent \textbf{Attention Concentration (Focus Degree)}

Maximum single-row attention weight:

\begin{equation}
m(\textit{a}) = \max_{k} a_k.
\end{equation}

Regional average concentration:
\begin{equation}
C(\mathcal{A}) = \frac{1}{N}\sum_{i} m(\mathcal{A}_{i,:}).
\end{equation}

\begin{algorithm}[!t]
\caption{VLALeaks: Two-Stage Membership Inference}
\label{alg:vla_leaks}
\begin{algorithmic}[1]
\STATE \textbf{Input}: Target VLA model $\mathcal{F}(\cdot)$, dataset with membership $\mathcal{X}=\{x_i\}_{i=1}^{n}$, attention layers ${L}$.
\STATE \textbf{Output}: Attack model $g_\phi$

\STATE \textbf{Stage 1: Membership Feature Extraction}
\FORALL{$x \in \mathcal{X}$ and $l \in L$}
    \STATE \textbf{Partitioning Rule} 
    \STATE Extract Attention
    $\{A^{(l)}_{{img}}, A^{(l)}_{{txt}}, A^{(l)}_{{act}}\}\leftarrow \mathcal{F}(x)$
    \STATE \textbf{Feature Modeling}
    \STATE $F_{\text{per-layer}}^{(l)} \leftarrow \text{per-layer}(\{A^{(l)}_{{img}}, A^{(l)}_{{txt}}, A^{(l)}_{{act}}\})$
    \STATE $F_{\text{cross-layer}}^{(l)} \leftarrow \text{cross-layer}(\{A^{(l)}_{{img}}, A^{(l)}_{{txt}}, A^{(l)}_{{act}}\})$
    \STATE $F_{\text{action}}^{(l)} \leftarrow \text{action}(\{A^{(l)}_{{img}}, A^{(l)}_{{txt}}, A^{(l)}_{{act}}\})$
    \STATE \textbf{Feature Fusion}
    \STATE $\phi(x) \leftarrow \text{concat}\big(F_{\text{per-layer}}^{(l)},F_{\text{cross-layer}}^{(l)},F_{\text{action}}^{(l)}\big)_{l\in{L}}$
\ENDFOR
\STATE Collect all features: $\Phi = \{\phi(x_i)\}_{i=1}^n$.

\STATE \textbf{Stage 2: Attack Model Construction}
\STATE Build membership inference attack feature set: $D_{\text{attack}} = \{(\phi(x_i), y_i)\}_{i=1}^n,\ y_i \in \{0,1\}$.
\STATE Initialize attack model $g_\phi$ and train on $D_{\text{attack}}$.

\RETURN $g_\phi$
\STATE \textbf{Inference}
\STATE Given a target sample $\boldsymbol{x}$, predict membership: $1/0 \leftarrow g_\phi(\phi(\boldsymbol{x}))$.

\end{algorithmic}
\end{algorithm}

(2) Cross-Layer Temporal Evolution Feature Extraction. Leveraging the per-layer feature extraction results across multiple layers, VLALeaks can uncover the evolution patterns of attention with network depth and construct inter-layer dynamic features, including:

\begin{itemize}
    \item Focusing trend features: The mean and fluctuation variance of cross-modal attention concentration across layers, reflecting the convergence trend of modality interaction focusing.
    \item Inter-layer similarity features: Frobenius norm distance and symmetric KL divergence are used to quantify the distribution differences of attention weights between adjacent layers, characterizing the stability of modality interaction patterns.
    \item Modality interaction propensity features: The difference between intra-modal interaction strength and cross-modal interaction strength is computed to determine whether the model prefers intra-modal associations or cross-modal semantic associations.
\end{itemize}

Characterize the dynamic variation of attention features with network depth.

\noindent \textbf{Attention Focus Trend}

Single-layer image-to-text attention concentration:
\begin{equation}
C^{(l)} = C\left(A_{{img2txt}}^{(l)}\right).
\end{equation}

Cross-layer concentration mean:
\begin{equation}
\bar{C} = \frac{1}{L}\sum_{l=0}^{L-1} C^{(l)}.
\end{equation}

Cross-layer concentration fluctuation variance:
\begin{equation}
\sigma_C = \sqrt{\frac{1}{L}\sum_{l=0}^{L-1}\left(C^{(l)}-\bar{C}\right)^2}.
\end{equation}
\noindent \textbf{Layer-wise Frobenius Distance}
\begin{equation}
d_F(P,Q) = \|P-Q\|_F=\sqrt{\sum_{i,j}(P_{ij}-Q_{ij})^2}.
\end{equation}

Adjacent-layer attention distribution distance:
\begin{equation}
d_F^{(l,l+1)} = d_F\left(A_{{img2txt}}^{(l)},A_{{img2txt}}^{(l+1)}\right).
\end{equation}
\noindent \textbf{Symmetric KL Divergence}

Basic KL divergence:
\begin{equation}
D_{\text{KL}}(P\parallel Q)=\sum_{k}P_k\log\frac{P_k+\varepsilon}{Q_k+\varepsilon}.
\end{equation}

Symmetric KL divergence:
\begin{equation}
D_{\text{SKL}}(P,Q)=\frac{1}{2}\big(D_{\text{KL}}(P\parallel Q)+D_{\text{KL}}(Q\parallel P)\big).
\end{equation}

Adjacent-layer distribution difference:
\begin{equation}
D_{\text{SKL}}^{(l,l+1)} = D_{\text{SKL}}\left(A_{{img2txt}}^{(l)},A_{{img2txt}}^{(l+1)}\right).
\end{equation}
\noindent \textbf{Intra-modal vs Cross-modal Interaction Tendency}
\begin{equation}
T^{(l)} = \frac{1}{2}\left[ \mu_{{img}}^{(l)}+\mu_{{txt}}^{(l)} -\mu_{{img2txt}}^{(l)}-\mu_{{txt2img}}^{(l)} \right].
\end{equation}

(3) Action-Modality-Specific Feature Extraction. An action attention extraction sub-module is invoked to supplement association features, covering bidirectional interaction logic:

\begin{itemize}
    \item Directed attention representations from action tokens to image and text modalities;
    \item Interaction features from image and text modalities back to action tokens;
    \item Self-attention association features within action tokens;
    \item Simultaneous statistics of action attention, including mean, variance, distribution entropy, and focusing degree, to complete the action semantic dimension representation.
\end{itemize}

\noindent \textbf{Action-to-Image \& Action-to-Text Attention Statistics}

Action-to-image attention mean:
\begin{equation}
\mu_{act2img}^{(l)} = \frac{1}{N_a\cdot S_{img}}\sum_{i,j}\mathcal{A}_{act2img,i,j}^{(l)}.
\end{equation}

Action-to-text attention mean:
\begin{equation}
\mu_{act2txt}^{(l)} = \frac{1}{N_a\cdot S_{txt}}\sum_{i,j}\mathcal{A}_{act2txt,i,j}^{(l)}.
\end{equation}

Action-to-image attention standard deviation:
\begin{equation}
\sigma_{act2img}^{(l)} = \sqrt{\frac{\sum_{i,j}\big(\mathcal{A}_{act2img,i,j}^{(l)}-\mu_{act2img}^{(l)}\big)^2}{N_a\cdot S_{img}}}.
\end{equation}

Action-to-text attention standard deviation:
\begin{equation}
\sigma_{act2txt}^{(l)} = \sqrt{\frac{\sum_{i,j}\big(\mathcal{A}_{act2txt,i,j}^{(l)}-\mu_{act2txt}^{(l)}\big)^2}{N_a\cdot S_{txt}}}.
\end{equation}

\noindent \textbf{Action Attention Entropy and Concentration}

Action region average entropy:
\begin{equation}
H_{act}^{(l)} = -\frac{1}{N_a}\sum_{i}\sum_{k}\mathcal{A}_{act,i,k}^{(l)}\log(\mathcal{A}_{act,i,k}^{(l)}+\varepsilon).
\end{equation}

Action region average concentration:
\begin{equation}
C_{act}^{(l)} = \frac{1}{N_a}\sum_{i}\max_{k}\mathcal{A}_{act,i,k}^{(l)}.
\end{equation}

\noindent \textbf{Image-to-Action \& Text-to-Action Attention}

Image-to-action attention mean:
\begin{equation}
    \mu_{img2act}^{(l)} = \frac{1}{S_{img}\cdot N_a}\sum_{i,j}\mathcal{A}_{img2act,i,j}^{(l)}.
\end{equation}

Text-to-action attention mean:
\begin{equation}
    \mu_{txt2act}^{(l)} = \frac{1}{S_{txt}\cdot N_a}\sum_{i,j}\mathcal{A}_{txt2act,i,j}^{(l)}.
\end{equation}

\noindent \textbf{Intra-action Self-Attention Features}

Intra-action attention mean:
\begin{equation}
    \mu_{act}^{(l)} = \frac{1}{N_a^2}\sum_{i,j}\mathcal{A}_{act,i,j}^{(l)}.
\end{equation}

Intra-action attention standard deviation:
\begin{equation}
    \sigma_{act}^{(l)} = \sqrt{\frac{1}{N_a^2}\sum_{i,j}\big(\mathcal{A}_{act,i,j}^{(l)}-\mu_{act}^{(l)}\big)^2}.
\end{equation}

\subsubsection{Feature Fusion}

The per-layer regional features, cross-layer evolution features, and action interaction features are uniformly encapsulated in a key-value structured format. A multi-dimensional attention feature dictionary aligned along the batch dimension is output. The feature dimensions cover four attribute categories: static regional interaction, dynamic cross-layer evolution, bidirectional cross-modal association, and action semantic binding. This achieves comprehensive extraction of multi-scale attention representations, thereby supporting the subsequent inference attack.

\subsection{Stage 2: Attack Model Construction}

In this stage, VLALeaks builds a binary classification attack model using the membership features extracted in Stage 1. This attack model is used to determine whether a given sample is a member sample. Given that all extracted features contribute membership information, we argue that leveraging all features improves attack accuracy. Details are provided in Section~\ref{iofn}. We adopt two lightweight models as attack models: Multilayer Perceptron (MLP)~\cite{taud2017multilayer}, and Random Forest (RF)~\cite{breiman2001random}. These models are well-suited for membership inference due to their ability to capture non-linear feature interactions (in the case of MLP) and their robustness to high-dimensional, heterogeneous feature spaces (in the case of RF), while maintaining low training and inference overhead. By employing both models, we also enable a comparative evaluation of different decision boundaries and ensemble behaviors in the context of membership leakage detection.

\section{Experiments}
\label{Experiments}
\subsection{Experimental Setup}

\subsubsection{Robot Setup}
\begin{figure}[!t]
    \centering
    \includegraphics[width=0.6\linewidth]{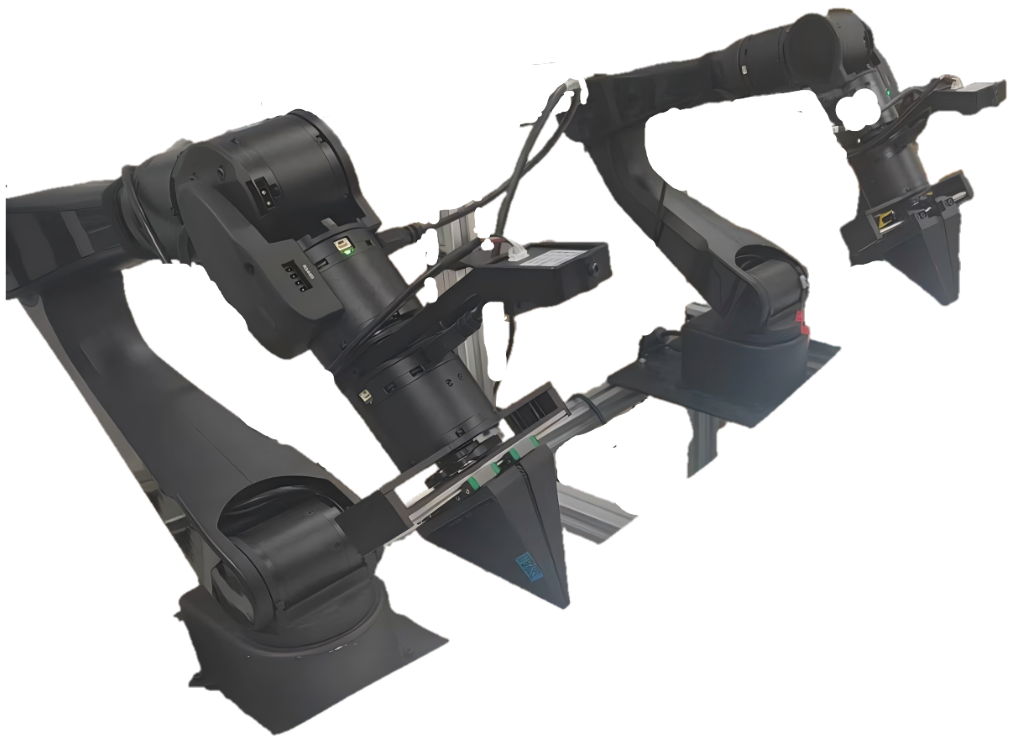}
    \caption{Real-world bimanual manipulation platform. Two AIRBOT Play 6-DOF arms, mounted on a fixed aluminum-profile frame, are each equipped with a parallel-jaw 1-DOF gripper and a wrist-mounted RGB camera. An overhead wide-angle camera is positioned above the midpoint between the two arms to capture the global workspace.}
    \label{fig:platform}
\end{figure}

For virtual simulation, we use the LIBERO platform~\cite{liu2023libero}, focusing on experiments involving OpenVLA~\cite{kim2024openvla} and OpenVLA-oft~\cite{kim2025fine}. For real-world robot evaluations, our platform adopts a symmetric bimanual setup consisting of two AIRBOT Play arms (see Figure~\ref{fig:platform}) as the execution (slave) units running $\pi_0$~\cite{black2024pi_0}, and two AIRBOT Replay arms serving as the teleoperation (master) units for data collection. Each AIRBOT Play arm is a 6-DOF serial manipulator with a parallel-jaw 1-DOF gripper (opening range: 0–60,mm), operating under joint position control. The joint motion limits are listed in Table~\ref{tab:joint_range}. Both arms are symmetrically mounted on a fixed aluminum-profile frame without a mobile base, creating a stationary bimanual workspace well-suited for tabletop manipulation tasks. During both teleoperation and autonomous execution, joint position commands are simultaneously streamed to both arms at 10\,Hz. The overall action space is 14-dimensional, consisting of six joint targets and one gripper command per arm.
\begin{table}[!t]
    \centering
    \caption{AIRBOT Play arm joint position ranges.}
    \label{tab:joint_range}
    \begin{tabular}{cc}
        \toprule
        \textbf{Joint} & \textbf{Range} \\
        \midrule
        J1 & $-180^\circ$ to $+120^\circ$ \\
        J2 & $-170^\circ$ to $+10^\circ$ \\
        J3 & $-5^\circ$ to $+180^\circ$ \\
        J4 & $-172.5^\circ$ to $+172.5^\circ$ \\
        J5 & $-105^\circ$ to $+105^\circ$ \\
        J6 & $-172^\circ$ to $+172^\circ$ \\
        \bottomrule
    \end{tabular}
\end{table}

The perception system consists of three RGB cameras providing complementary viewpoints:

\begin{itemize}
    \item \textbf{Wrist cameras ($\times$2).} Each arm is equipped with a wrist-mounted RGB camera (focal length: 2.8\,mm, field of view: 75$^\circ$, minimal distortion) that captures 720p (1280$\times$720) images. These cameras provide close-range views of the grasping region, offering fine-grained visual cues for object localization and gripper alignment.
    \item \textbf{Overhead camera ($\times$1).} A wide-angle RGB camera (focal length: 2.1\,mm, 140$^\circ$ FoV) is mounted above and between the two arms, pointing downward to capture a global overhead view of the entire workspace. The short focal length introduces noticeable barrel distortion, which is corrected through calibrated undistortion before policy input. This wide field of view provides the full spatial layout of all trays and objects, allowing the policy to reason about tray selection and coordinate bimanual actions.
\end{itemize}
All three cameras synchronously stream RGB images at 720p resolution during both data collection and policy inference. Their specifications are summarized in Table~\ref{tab:camera_spec}.

\begin{table}[!t]
    \centering
    \caption{Camera configurations for the bimanual platform.}
    \label{tab:camera_spec}
    \adjustbox{width=\linewidth}{
    \begin{tabular}{lccc}
        \toprule
        \textbf{Camera} & \textbf{Resolution} & \textbf{Focal Length / FoV} & \textbf{Distortion} \\
        \midrule
        Wrist ($\times$2) & $1280 \times 720$ & 2.8\,mm / 75$^\circ$ & Negligible \\
        Overhead ($\times$1) & $1280 \times 720$ & 2.1\,mm / 140$^\circ$ & Barrel (corrected) \\
        \bottomrule
    \end{tabular}
    }
\end{table}

\subsubsection{Dataset}
\label{ddd}
For OpenVLA~\cite{kim2024openvla} and OpenVLA-oft~\cite{kim2025fine}, we perform MIAs and evaluation using the LIBERO dataset~\cite{liu2023libero} (virtual simulation).
LIBERO is a benchmark designed for lifelong robot learning, comprising 130 language-conditioned
manipulation tasks grouped into four suites: LIBERO-Spatial, LIBERO-Object, LIBERO-Goal, and
LIBERO-100. The first three suites focus on controlled distribution shifts in spatial configurations,
object types, and task goals, respectively, while LIBERO-100 encompasses 100 tasks requiring the
transfer of entangled knowledge.

For $\pi_0$~\cite{black2024pi_0}, we collect real-world VLA data from physical robot experiments. We evaluate our method on a bimanual tabletop manipulation benchmark set in a grocery store scenario. The workspace consists of a table with multiple trays arranged in sequence, each containing items of a single category. The robot must visually identify the target tray and grasp the specified object using coordinated bimanual control. We define six tasks (a)-(f) as follows:

\begin{itemize}
    \item (a) \textbf{Pull the tray right.} Moving the tray to the right is considered a success.
    \item (b) \textbf{Pick up a chocolate and put it in the tray.} The robot identifies the tray containing chocolate boxes among several trays on the table and grasps one box from it. A trial is considered successful if the gripper firmly holds the chocolate box and lifts it clear of the tray. Starting with an object already held in the gripper, the robot places it into a designated black tray. The trial is successful if the object is released inside the tray and remains stable after placement.
    \item (c) \textbf{Pick up a cola and put it in the tray.} The robot identifies the tray containing cola bottles and picks up one bottle. Success is defined as the bottle being firmly grasped and lifted without slipping. Starting with an object already held in the gripper, the robot places it into a designated black tray. The trial is successful if the object is released inside the tray and remains stable after placement.
    \item (d) \textbf{Pick up a candy and put it back.} The robot locates the tray holding mint candies and picks up one piece. Success requires the candy to be securely grasped and lifted without dropping, and then placed back.
    \item (e) \textbf{Pull the tray left.} Moving the tray to the left is considered a success.
    \item (f) \textbf{Pick up a snicker and put it in the tray.} The robot finds the tray with Snickers bars and grasps one bar. The trial succeeds when the bar is stably held above the tray surface. Starting with an object already held in the gripper, the robot places it into a designated black tray. The trial is successful if the object is released inside the tray and remains stable after placement.
\end{itemize}






We adopt a continuous collection strategy to maximize data diversity within each task category. For a given tray containing $\mathbb{N}$ instances of the same item (e.g., five mint candies), the teleoperator performs $\mathbb{N}$ consecutive grasping episodes without resetting the scene, naturally producing variations in object arrangement, grasp pose, and relative positioning as the tray becomes progressively emptier. This sequential depletion protocol yields diverse demonstrations without manual scene randomization. Each task category contains a minimum of 20 demonstrated trajectories. All demonstrations are collected via bilateral teleoperation at 10\,Hz used during policy execution.

\begin{table*}
  \centering
  \caption{Experimental results on OpenVLA.}
  \label{tab:openvla}
  {
  \begin{tabular}{lcccccccc}
    \toprule
    & \multicolumn{2}{c}{LIBERO-Spatial} & \multicolumn{2}{c}{LIBERO-Object} & \multicolumn{2}{c}{LIBERO-Goal} & \multicolumn{2}{c}{LIBERO-10} \\
    \cmidrule(lr){2-3} \cmidrule(lr){4-5} \cmidrule(lr){6-7} \cmidrule(lr){8-9}
    Method & AUC & TPR@1\%FPR & AUC & TPR@1\%FPR & AUC & TPR@1\%FPR & AUC & TPR@1\%FPR \\
    \midrule
    Loss       & 0.5166 & 0.0152 & 0.4658 & 0.0052 & 0.4575 & 0.0016 & 0.4920 & 0.0104 \\
    Min-k      & 0.5980 & 0.0256 & 0.6084 & 0.0352 & 0.5319 & 0.0452 & 0.5217 & 0.0088 \\
    Min-k++    & 0.5886 & 0.0272 & 0.5607 & 0.0268 & 0.6181 & 0.0212 & 0.5828 & 0.0236 \\
    MaxR-img   & 0.5504 & 0.0170 & 0.4984 & 0.0276 & 0.8155 & 0.1409 & 0.6267 & 0.0294 \\
    MaxR-inst  & 0.5363 & 0.0150 & 0.4917 & 0.0070 & 0.6015 & 0.0502 & 0.7031 & 0.0484 \\
    MaxR-desp  & 0.4203 & 0.0072 & 0.4837 & 0.0082 & 0.4821 & 0.0093 & 0.4920 & 0.0104 \\
    SMI        & 0.9515 & 0.3691 & 0.6970 & 0.1008 & 0.5501 & 0.1410 & 0.8699 & 0.1936 \\
    RIM        & 0.9817 & 0.7520 & 0.9222 & 0.4210 & 0.8716 & 0.2740 & 0.9393 & 0.4520 \\
    RINM       & 0.8827 & 0.2560 & 0.7580 & 0.1190 & 0.7629 & 0.1070 & 0.8877 & 0.2520 \\
    \midrule
    \rowcolor{gray!30}
    VLALeaks-RF & \multicolumn{1}{c}{\bfseries 0.9946} & \multicolumn{1}{c}{\bfseries 0.8910} & \multicolumn{1}{c}{\bfseries 0.9940} & \multicolumn{1}{c}{\bfseries 0.8758} & \multicolumn{1}{c}{\bfseries 0.9508} & \multicolumn{1}{c}{\bfseries 0.4430} & \multicolumn{1}{c}{\bfseries 0.9809} & \multicolumn{1}{c}{\bfseries 0.7563} \\
    \rowcolor{gray!30}
    VLALeaks-MLP& \multicolumn{1}{c}{\bfseries 0.9991} & \multicolumn{1}{c}{\bfseries 0.9774} & \multicolumn{1}{c}{\bfseries 0.9987} & \multicolumn{1}{c}{\bfseries 0.9798} & \multicolumn{1}{c}{\bfseries 0.9300} & \multicolumn{1}{c}{\bfseries 0.3406} & \multicolumn{1}{c}{\bfseries 0.9959} & \multicolumn{1}{c}{\bfseries 0.9310} \\
    \bottomrule
  \end{tabular}
  }
\end{table*}

\begin{table*}
  \centering
  \caption{Experimental results on OpenVLA-oft.}
  \label{tab:openvla_oft}
  {
  \begin{tabular}{lcccccccc}
    \toprule
    & \multicolumn{2}{c}{LIBERO-Spatial} & \multicolumn{2}{c}{LIBERO-Object} & \multicolumn{2}{c}{LIBERO-Goal} & \multicolumn{2}{c}{LIBERO-10} \\
    \cmidrule(lr){2-3} \cmidrule(lr){4-5} \cmidrule(lr){6-7} \cmidrule(lr){8-9}
    Method & AUC & TPR@1\%FPR & AUC & TPR@1\%FPR & AUC & TPR@1\%FPR & AUC & TPR@1\%FPR \\
    \midrule
    Loss       & 0.5042 & 0.0132 & 0.4979 & 0.0122 & 0.5369 & 0.0114 & 0.4892 & 0.0078 \\
    Min-k      & 0.4769 & 0.0070 & 0.4248 & 0.0124 & 0.4522 & 0.0104 & 0.3593 & 0.0060 \\
    Min-k++    & 0.5028 & 0.0128 & 0.4229 & 0.0098 & 0.5295 & 0.0174 & 0.3358 & 0.0030 \\
    MaxR-img   & 0.5300 & 0.0170 & 0.6001 & 0.0238 & 0.4623 & 0.0048 & 0.4810 & 0.0092 \\
    MaxR-inst  & 0.5225 & 0.0176 & 0.4458 & 0.0134 & 0.4626 & 0.0038 & 0.3729 & 0.0030 \\
    MaxR-desp  & 0.4723 & 0.0056 & 0.4111 & 0.0066 & 0.4690 & 0.0090 & 0.3911 & 0.0038 \\
    SMI        & 0.9085 & 0.4707 & 0.9774 & 0.6020 & 0.9704 & 0.6024 & 0.9692 & 0.6193 \\
    RIM        & 0.6331 & 0.0240 & 0.7159 & 0.0780 & 0.6624 & 0.0270 & 0.6936 & 0.0560 \\
    RINM       & 0.5737 & 0.0160 & 0.8955 & 0.2840 & 0.6094 & 0.0220 & 0.7049 & 0.0510 \\
    \midrule
    \rowcolor{gray!30}
    VLALeaks-RF & \multicolumn{1}{c}{\bfseries 0.9857} & \multicolumn{1}{c}{\bfseries 0.7742} & \multicolumn{1}{c}{\bfseries 0.9950} & \multicolumn{1}{c}{\bfseries 0.9106} & \multicolumn{1}{c}{\bfseries 0.9896} & \multicolumn{1}{c}{\bfseries 0.8099} & \multicolumn{1}{c}{\bfseries 0.9930} & \multicolumn{1}{c}{\bfseries 0.9094} \\
    \rowcolor{gray!30}
    VLALeaks-MLP& \multicolumn{1}{c}{\bfseries 0.9980} & \multicolumn{1}{c}{\bfseries 0.9597} & \multicolumn{1}{c}{\bfseries 0.9995} & \multicolumn{1}{c}{\bfseries 0.9921} & \multicolumn{1}{c}{\bfseries 0.9976} & \multicolumn{1}{c}{\bfseries 0.9493} & \multicolumn{1}{c}{\bfseries 0.9972} & \multicolumn{1}{c}{\bfseries 0.9501} \\
    \bottomrule
  \end{tabular}
  }
\end{table*}

\subsubsection{Evaluation Metrics}
Following prior works~\cite{hu2025membership,li2024membership,zhang2025min}, we adopt two standard metrics for membership inference: Attack AUC, which quantifies the method's overall discriminative ability, and TPR@1\%FPR, which captures performance in the low-false-positive regime critical for MIAs. A high TPR@1\%FPR indicates that training samples can be identified with high confidence while incurring few false alarms. All metrics are reported as averages across cross-validation folds.

\subsubsection{Baselines}
We compare our proposed VLALeaks against nine state-of-the-art MIAs. Loss attack~\cite{yeom2018privacy} exploits differences in loss values on machine learning models. Min-k~\cite{shi2024detecting} and Min-k++~\cite{zhang2025min} leverage predicted token probabilities to perform text-modal MIAs on LLMs. MaxRényi~\cite{li2024membership} computes the Rényi entropy of predicted token probabilities and applies attacks to either the text or image modality on VLMs, with three variants evaluated in our study: img, inst, and desp. Additionally, Shadow Model Inference (SMI)~\cite{hu2025membership}, Reference Inference with Member (RIM)~\cite{hu2025membership}, and Reference Inference with Non-Member (RINM)~\cite{hu2025membership} exploit temperature sensitivity to conduct MIAs on image-text pairs on VLMs.

We employ nine of the most representative MIAs as baselines to demonstrate the superiority of our proposed VLALeaks. We adhere to the original experimental settings for the baselines and evaluate the target VLA models.

\subsection{Main Results}

\subsubsection{Attack AUC and TPR@1\%FPR}
\label{attackauc}
As summarized in Table~\ref{tab:openvla} and Table~\ref{tab:openvla_oft}, our proposed VLALeaks (including variants based on RF and MLP) consistently outperforms all baseline MIAs across all four LIBERO task splits (LIBERO-Spatial, LIBERO-Object, LIBERO-Goal, LIBERO-100) on both vanilla OpenVLA and OpenVLA-oft in terms of AUC and TPR@1\%FPR. For the OpenVLA model, VLALeaks-MLP achieves near-perfect AUC values exceeding 0.99 on LIBERO-Spatial, LIBERO-Object, and LIBERO-100, with TPR@1\%FPR higher than 0.93 across these three subsets, substantially surpassing previous state-of-the-art RIM (0.7520, 0.4210, and 0.4520, respectively). Notably, LIBERO-Goal serves as the most challenging setting where all baseline methods suffer severe performance degradation; even under this tough scenario, both VLALeaks-RF and VLALeaks-MLP retain obvious superiority against existing competitors. When transferred to OpenVLA-oft, conventional attacks (e.g., RIM) experience a dramatic performance drop, while our VLALeaks maintains robust attacking efficacy: VLALeaks-MLP obtains TPR@1\%FPR above 0.94 over all four dataset partitions, demonstrating exceptional resilience. Further ablation between two instantiations reveals that the MLP variant generally surpasses the RF counterpart by modeling non-linear latent privacy patterns hidden inside VLA models, except for the vanilla LIBERO-Goal case, where RF achieves marginally better TPR@1\%FPR. Overall, comprehensive comparisons validate the superior effectiveness, generalization, and robustness of VLALeaks for membership inference against VLA models.

\begin{figure*}[t]
    \centering
    \begin{subfigure}[b]{0.49\linewidth}
        \centering
        \includegraphics[width=\linewidth]{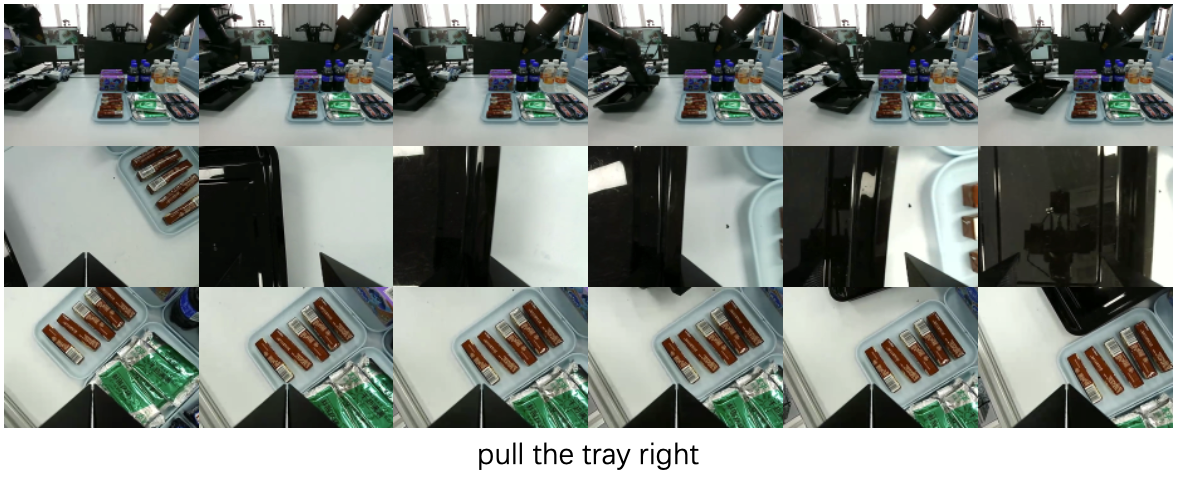}
        \caption{VLALeaks-RF(AUC=0.99) VLALeaks-MLP(AUC=0.98)}
        \label{fig:1}
    \end{subfigure}
    \hfill
    \begin{subfigure}[b]{0.49\linewidth}
        \centering
        \includegraphics[width=\linewidth]{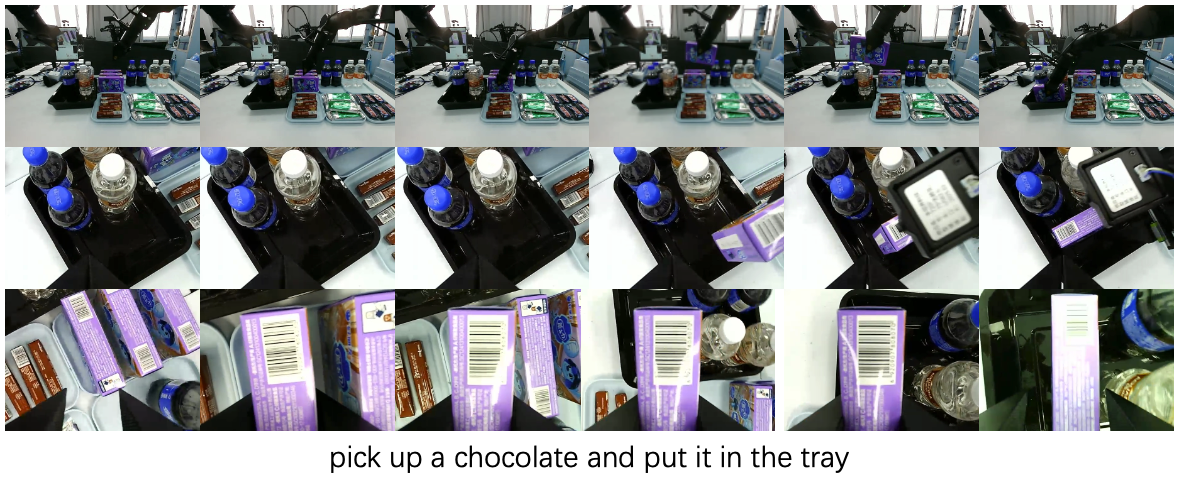}
        \caption{VLALeaks-RF(AUC=0.98) VLALeaks-MLP(AUC=0.97)}
        \label{fig:2}
    \end{subfigure}

    \vspace{1em}

    \begin{subfigure}[b]{0.49\linewidth}
        \centering
         \includegraphics[width=\linewidth]{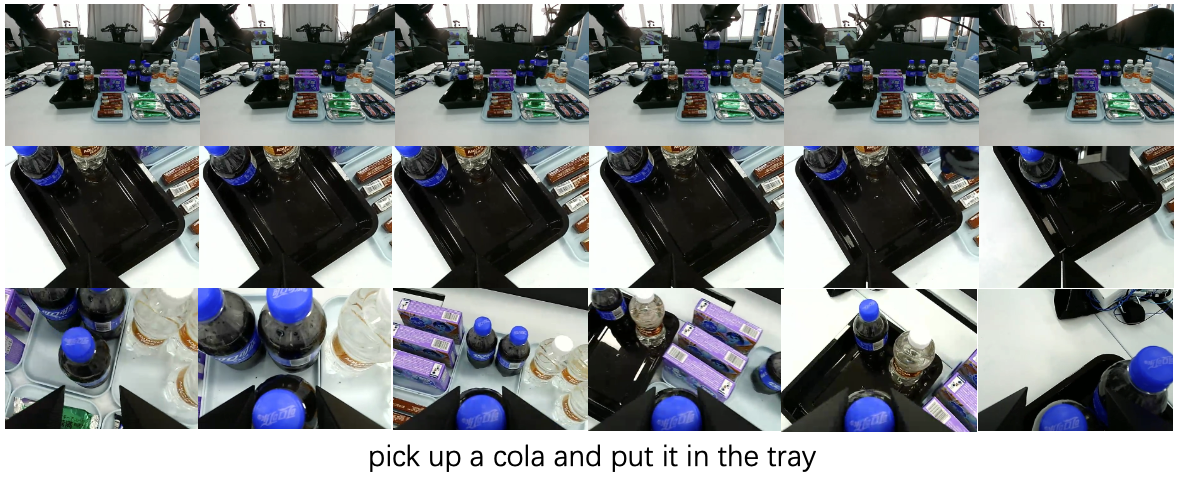}
        \caption{VLALeaks-RF(AUC=0.98) VLALeaks-MLP(AUC=0.99)}
        \label{fig:3}
    \end{subfigure}
    \hfill
    \begin{subfigure}[b]{0.49\linewidth}
        \centering
        \includegraphics[width=\linewidth]{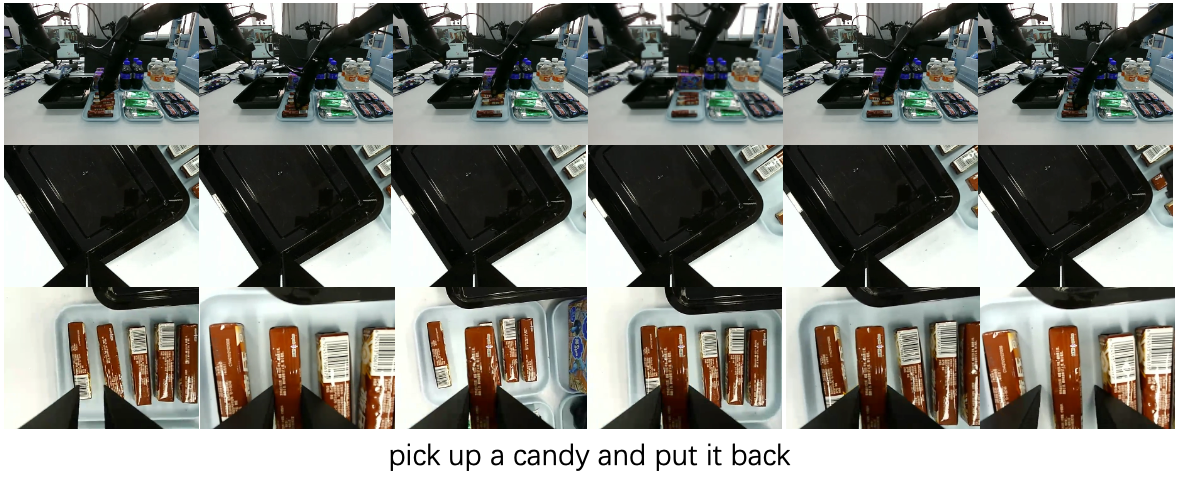}
        \caption{VLALeaks-RF(AUC=0.95) VLALeaks-MLP(AUC=0.96)}
        \label{fig:4}
    \end{subfigure}

    \vspace{1em}

    \begin{subfigure}[b]{0.49\linewidth}
        \centering
        \includegraphics[width=\linewidth]{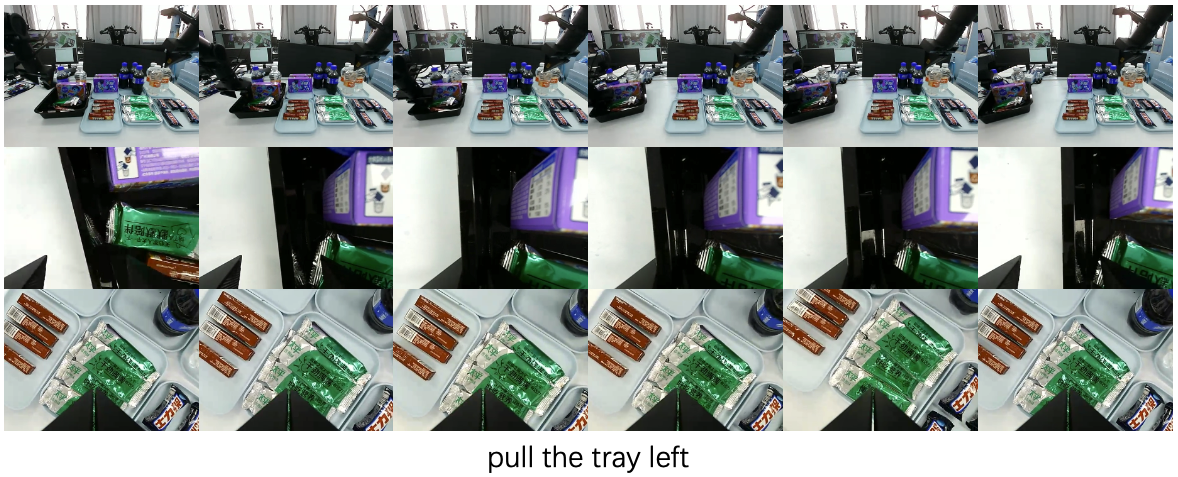}
        \caption{VLALeaks-RF(AUC=0.97) VLALeaks-MLP(AUC=0.98)}
        \label{fig:5}
    \end{subfigure}
    \hfill
    \begin{subfigure}[b]{0.49\linewidth}
        \centering
        \includegraphics[width=\linewidth]{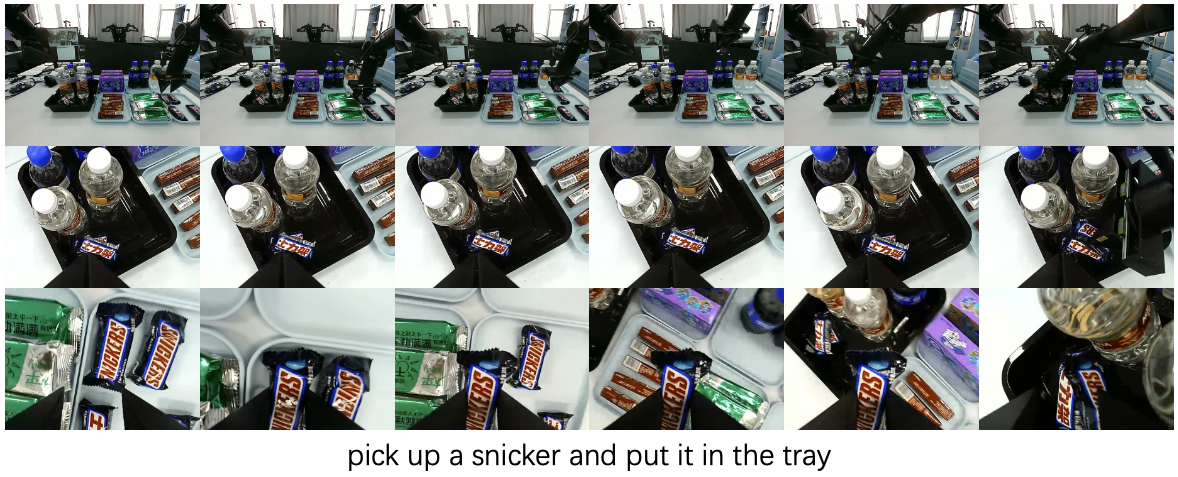}
        \caption{VLALeaks-RF(AUC=0.98) VLALeaks-MLP(AUC=0.98)}
        \label{fig:6}
    \end{subfigure}

    \caption{Attack AUC of VLALeaks on $\pi_0$ across six real-world tasks.}
    \label{fig:VLApi0}
\end{figure*}

We further conduct real-world experiments across six distinct robotic manipulation tasks in Figure~\ref{fig:VLApi0} to intuitively verify the attack performance of VLALeaks on VLA models. As illustrated in subfigures (a)–(f) covering diverse manipulation operations (tray pulling, object picking-and-placing for chocolates, cola, candies, and snickers), both VLALeaks-RF and VLALeaks-MLP achieve consistently high attack AUC scores ranging from 0.95 to 0.99 on individual downstream tasks, which demonstrates that our proposed VLALeaks can reliably capture implicit member privacy information embedded in multi-step robot trajectory data across varied manipulation paradigms.
Specifically, RF and MLP variants deliver complementary performance: VLALeaks-RF outperforms MLP on tasks (a)(b) with AUC=0.99 vs 0.98 and 0.98 vs 0.97, whereas VLALeaks-MLP obtains superior AUC results on task (c)(d) (0.99 vs 0.98, 0.96 vs 0.95). For remaining tasks (e)(f), RF and MLP yield nearly identical AUC values of ~0.97–0.98, further confirming the stable and robust attack capability of VLALeaks regardless of task-specific action logic differences.
Overall, the per-task visualization results align well with our quantitative benchmark results in Tables~\ref{tab:openvla}, ~\ref{tab:openvla_oft}, solidifying that VLALeaks effectively exposes membership privacy risks of VLA models across real-world robot manipulation scenarios. Additional experimental results are provided in Appendix~\ref{AdditionalExperiment}.

\subsubsection{PCA and t-SNE Visualization}
\begin{figure*}[]
    \centering
    \begin{subfigure}[b]{0.49\linewidth}
        \centering
        \includegraphics[width=\linewidth]{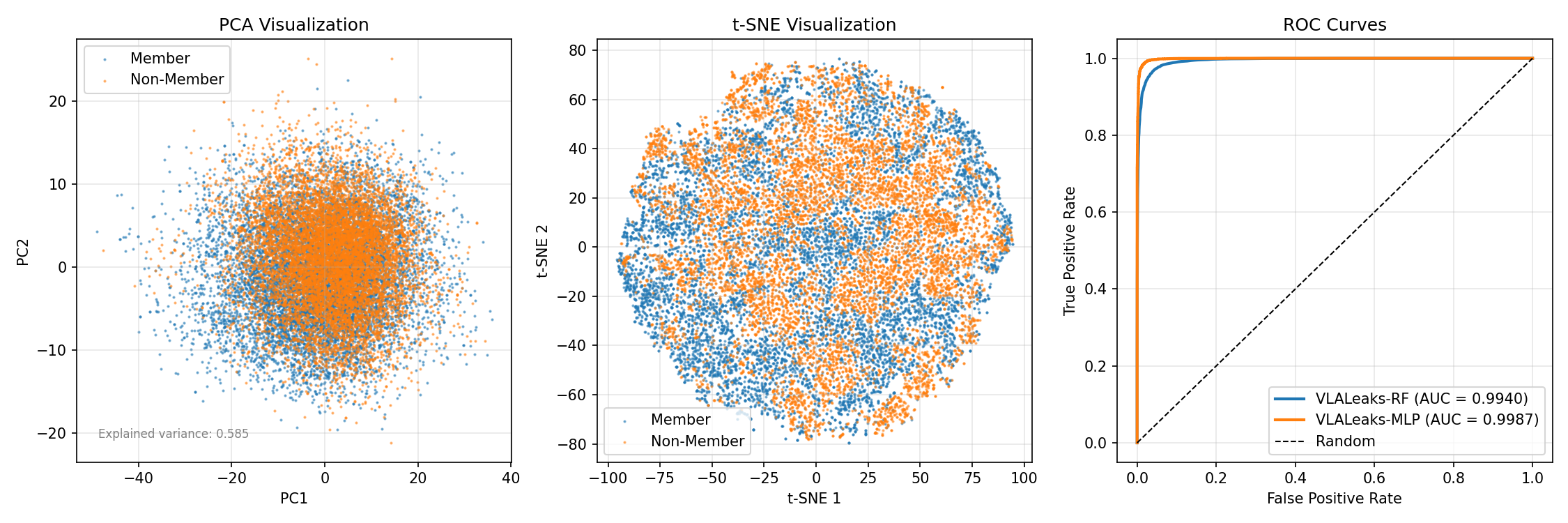}
        \caption{LIBERO-Spatial}
        \label{fig:feature_analysis_with_auc_spa}
    \end{subfigure}
    \hfill
    \begin{subfigure}[b]{0.49\linewidth}
        \centering
        \includegraphics[width=\linewidth]{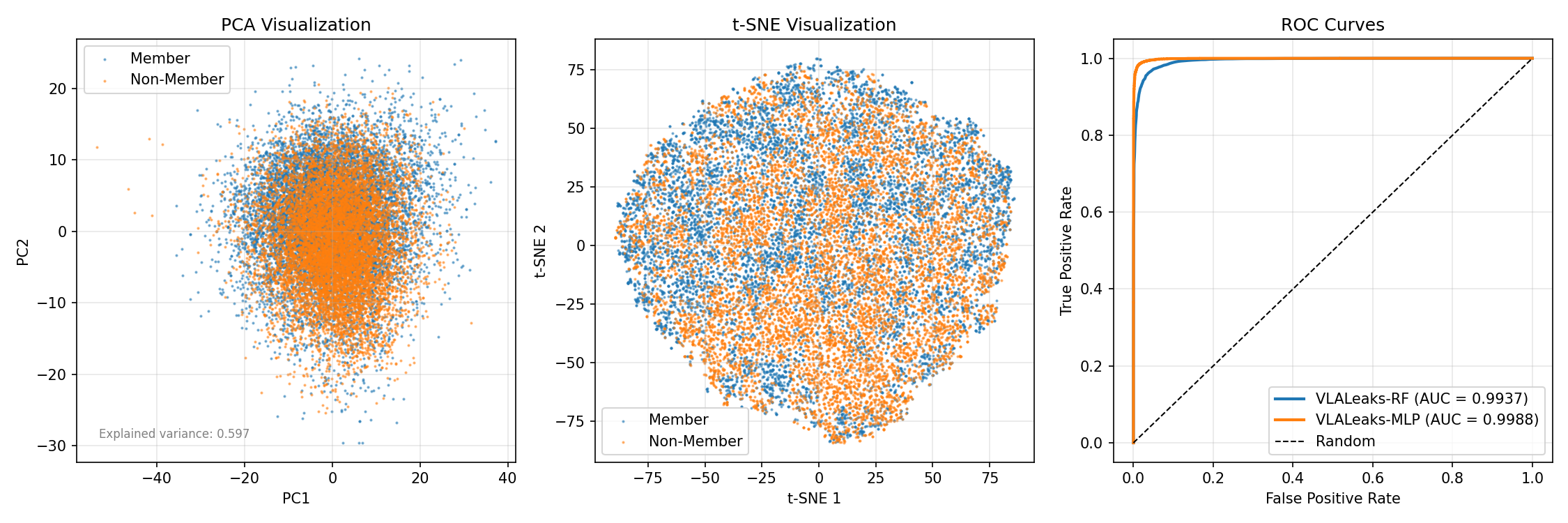}
        \caption{LIBERO-Object}
        \label{fig:feature_analysis_with_auc_obj}
    \end{subfigure}

    \vspace{1em} 

    \begin{subfigure}[b]{0.49\linewidth}
        \centering
        \includegraphics[width=\linewidth]{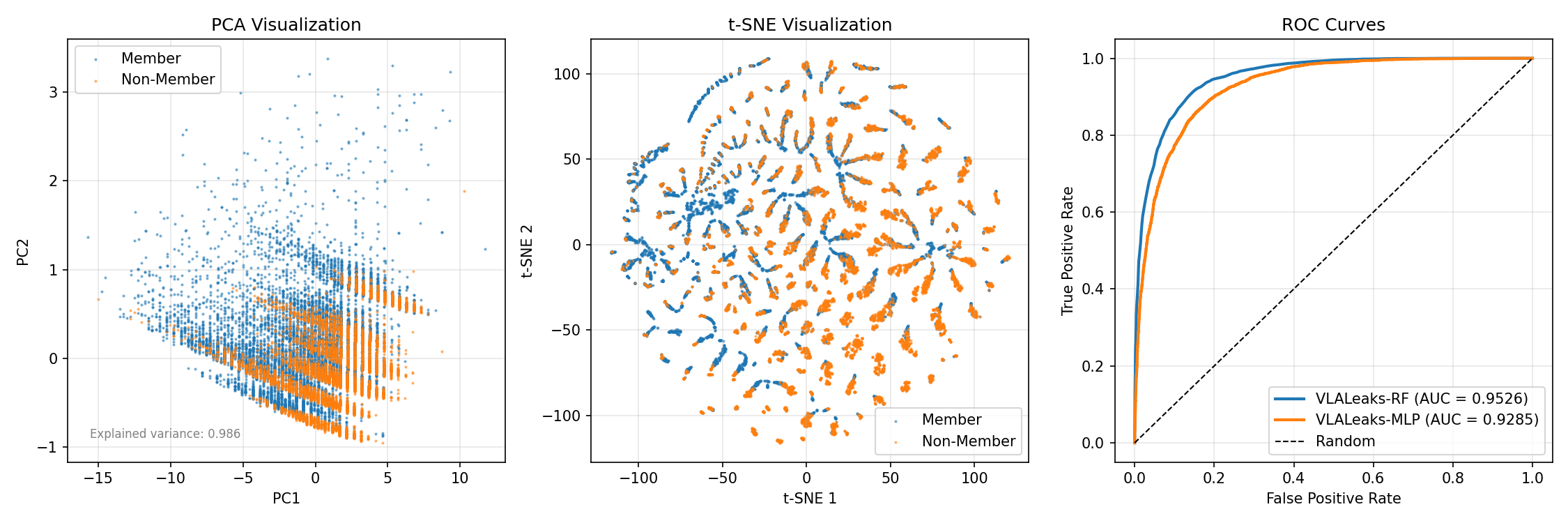}
        \caption{LIBERO-Goal}
        \label{fig:feature_analysis_with_auc_goal}
    \end{subfigure}
    \hfill
    \begin{subfigure}[b]{0.49\linewidth}
        \centering
        \includegraphics[width=\linewidth]{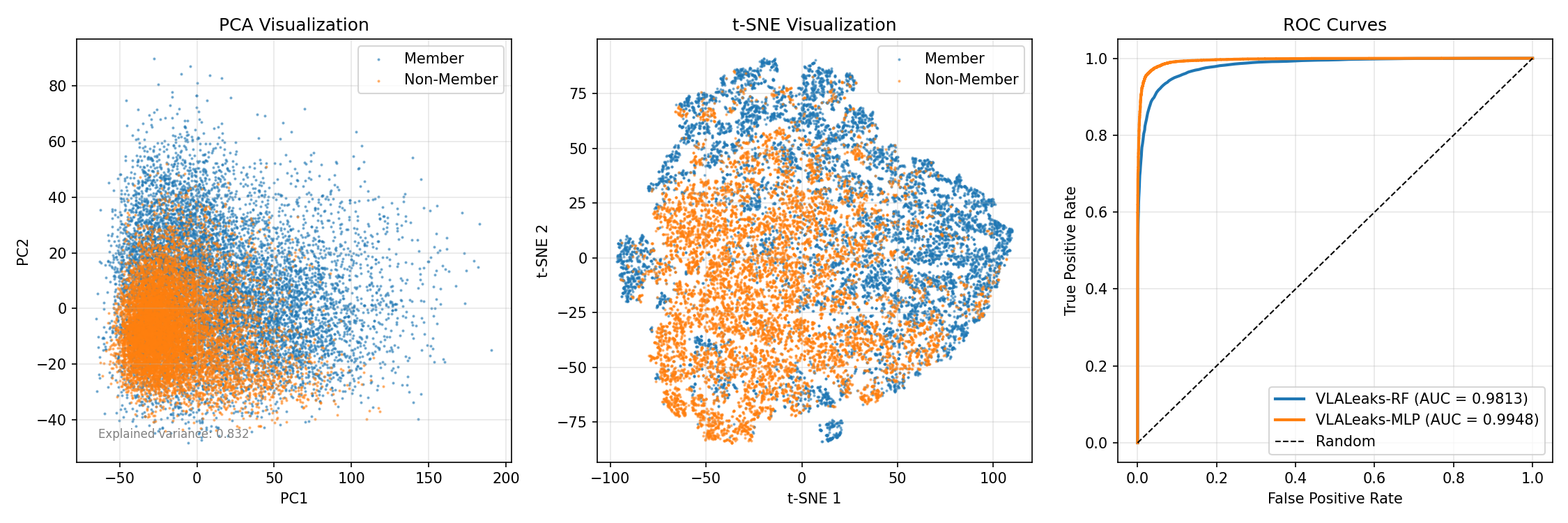}
        \caption{LIBERO-100}
        \label{fig:feature_analysis_with_auc_10}
    \end{subfigure}

    \caption{PCA and t-SNE of member and non-member features under OpenVLA on four datasets. PCA and t-SNE feature dimensionality reduction visualization of member and non-member feature distributions under OpenVLA on four LIBERO robotic datasets. PCA is utilized for linear dimensionality reduction to extract core feature components and eliminate redundant information, while t-SNE achieves nonlinear dimensionality reduction, intuitively visualizing the high-dimensional feature clustering and separation differences between member and non-member features in low-dimensional space.}
    \label{fig:pcatsne}
\end{figure*}

To further interpret the underlying mechanism of VLALeaks, we visualize the extracted privacy features of member (blue) and non-member (orange) samples via Principal Component Analysis (PCA)~\cite{abdi2010principal} and t-distributed Stochastic Neighbor Embedding (t-SNE)~\cite{van2008visualizing} dimensionality reduction alongside corresponding ROC curves across four LIBERO task splits in Figure~\ref{fig:pcatsne}.
From the feature distribution perspective, PCA and t-SNE plots consistently demonstrate obvious separation boundaries between member and non-member representations on LIBERO-Spatial, LIBERO-Object and LIBERO-100 (subfigure a/b/d): the two categories of samples form concentrated, mutually isolated clusters rather than chaotic overlapping distributions. Such distinct feature discrepancy fundamentally explains why VLALeaks-RF and VLALeaks-MLP achieve ultra-high AUC values above 0.98 in these three easy-to-discriminate scenarios, as the learned privacy embedding naturally encodes discriminative membership clues from VLA model outputs.
For the most challenging LIBERO-Goal dataset (subfigure c), member and non-member points partially overlap in low-dimensional projections, which accounts for the relatively declined AUC scores (0.9526 for RF, 0.9285 for MLP) compared with other three tasks. Nevertheless, VLALeaks still maintains competitive attack performance with AUC over 0.92, verifying our method can excavate subtle implicit membership differences even when privacy features are heavily entangled under complex goal-conditioned robot manipulation tasks.
In terms of ROC curves, both RF and MLP variants produce curves sharply close to the top-left corner across all four benchmarks and drastically outperform the random-guess baseline (diagonal dashed line). Meanwhile, the complementary performance between two backbones is observed: MLP gains marginal superiority on Spatial/Object/100 thanks to its powerful non-linear fitting capacity for continuous feature space, while RF performs slightly better on the tough LIBERO-Goal split by capturing discrete statistical privacy patterns.
Overall, the dimensionality reduction visualization intuitively reveals that VLALeaks effectively disentangles latent membership information hidden inside VLA’s vision-action features, which provides solid feature-level evidence for the outstanding quantitative results reported in previous tables.

\subsubsection{Impact of Features Number}
\label{iofn}
\begin{figure}[!t]
    \centering
    \begin{subfigure}[b]{\linewidth}
        \centering
        \includegraphics[width=\linewidth]{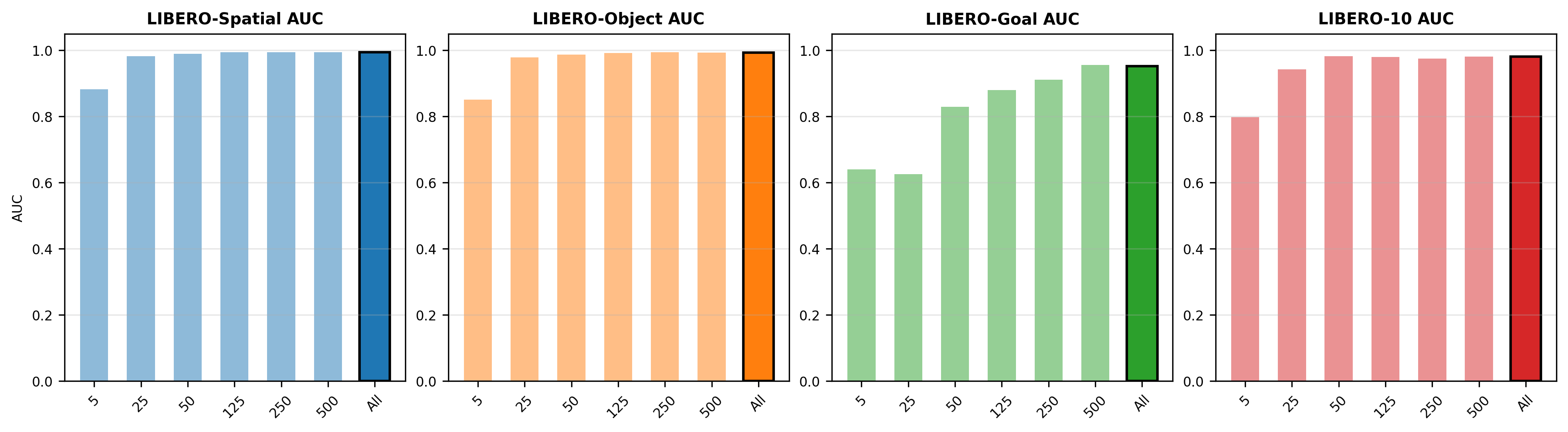}
        \caption{Attack AUC}
        \label{fig:rf_auc}
    \end{subfigure}
    \\[10pt]  
    \begin{subfigure}[b]{\linewidth}
        \centering
        \includegraphics[width=\linewidth]{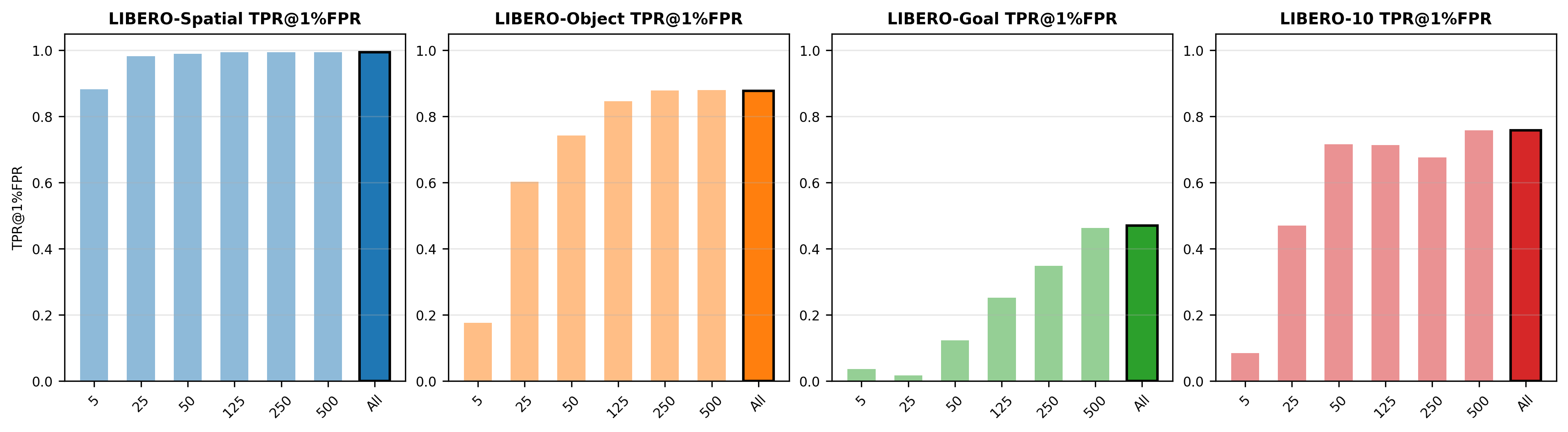}
        \caption{TPR@1\%FPR}
        \label{fig:rf_tpr}
    \end{subfigure}
    \caption{Attack Performance of VLALeaks-RF using varying numbers of features.}
    \label{fig:rf_combined}
\end{figure}

\begin{figure}[!t]
    \centering
    \begin{subfigure}[b]{\linewidth}
        \centering
        \includegraphics[width=\linewidth]{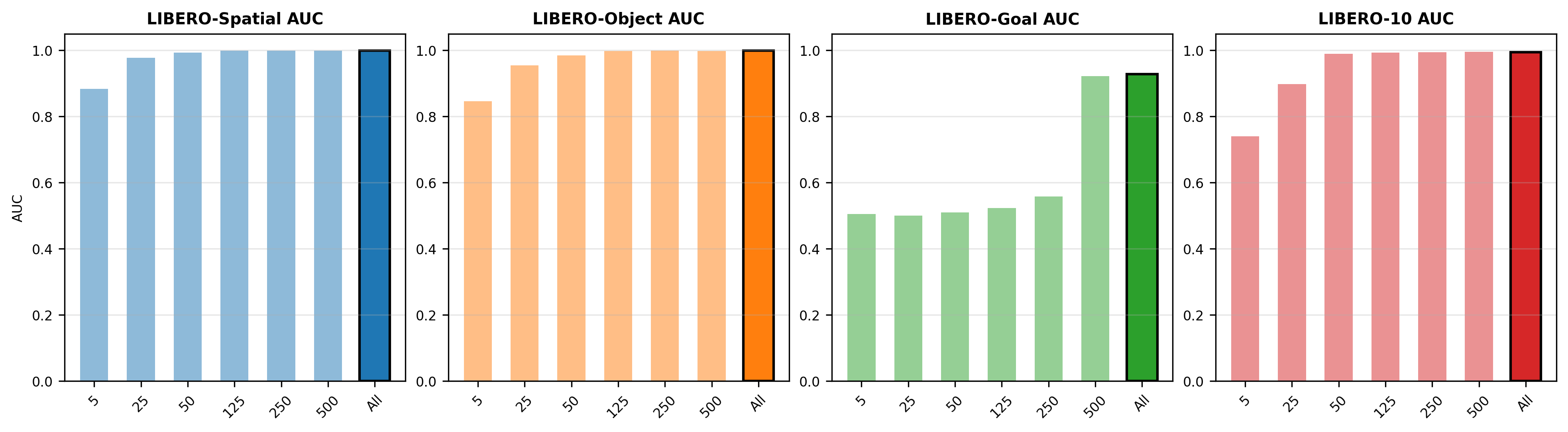}
        \caption{Attack AUC}
        \label{fig:mlp_auc}
    \end{subfigure}
    \\[10pt]  
    \begin{subfigure}[b]{\linewidth}
        \centering
        \includegraphics[width=\linewidth]{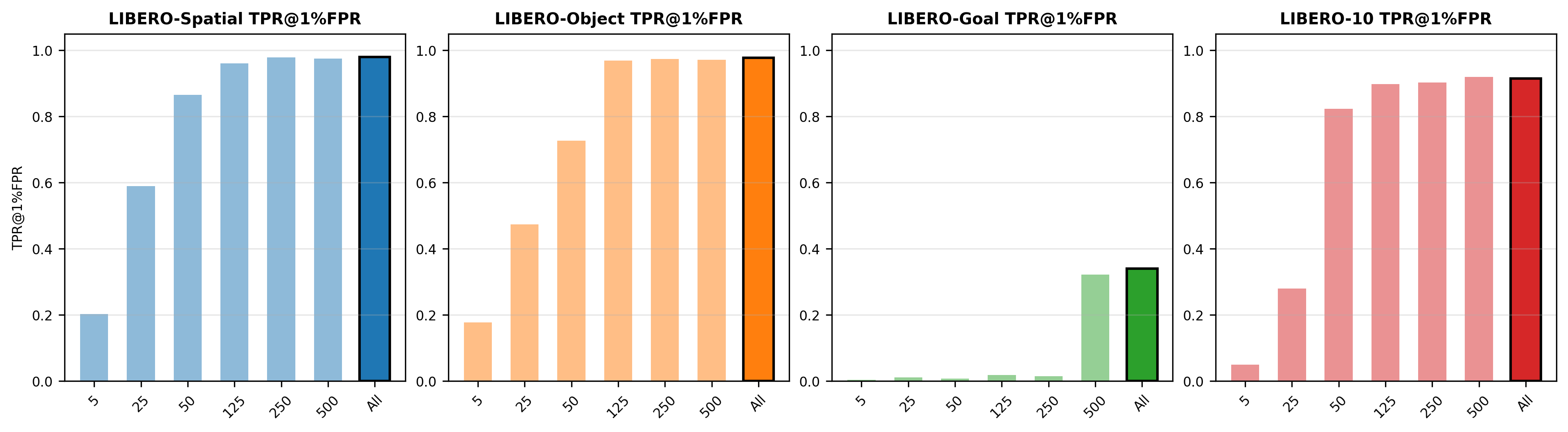}
        \caption{TPR@1\%FPR}
        \label{fig:mlp_tpr}
    \end{subfigure}
    \caption{Attack Performance of VLALeaks-MLP using varying numbers of features.}
    \label{fig:mlp_combined}
\end{figure}

We perform feature-size ablation studies in Figure~\ref{fig:rf_combined} (VLALeaks-RF) and Figure~\ref{fig:mlp_combined} (VLALeaks-MLP), gradually increasing the input feature dimension from 5, 25, 50, 125, 250, 500 to the full feature set (All) across four LIBERO tasks, with AUC and critical TPR@1\%FPR as evaluation metrics. For both RF and MLP variants on all four datasets, attack performance consistently rises as the number of selected features increases. Even with only the top-5 most informative features, VLALeaks already achieves a decent baseline AUC above 0.6 across all tasks; when feeding all extracted privacy features (All), both models reach their peak AUC and TPR@1\%FPR. This trend demonstrates that the privacy clues hidden inside VLA models distribute across the full feature space, and our feature selection module can effectively rank high-value privacy-contributing features sequentially.

On LIBERO-Spatial, LIBERO-Object and LIBERO-100, both RF and MLP converge rapidly: performance saturates early at 250-500 features, and full features only bring marginal extra improvement, with final AUC approaching 0.98-0.99. In particular, TPR@1\%FPR climbs steeply from extremely low values under sparse-feature settings to near 0.95 with full features, verifying abundant discriminative privacy information exists for these three scenarios.
LIBERO-Goal exhibits distinctly slower performance gain: under small feature counts (5/25/50), AUC and TPR@1\%FPR remain substantially lower than other subsets, and performance continuously climbs even up to full feature input without obvious saturation. It indicates membership privacy is sparsely encoded and scattered in feature dimensions for complex long-horizon goal tasks, requiring sufficient feature volume to excavate subtle member-nonmember discrepancies.

For low-feature regimes (5/25/50 features): RF generally outperforms MLP, since random forest excels at mining discrete statistical privacy from limited handpicked features.
For high-feature and full-feature input: MLP catches up and exceeds RF, benefiting from its strong non-linear fitting ability to model complex latent correlations inside dense high-dimensional privacy features, which aligns with our prior quantitative benchmark results.

Overall, the ablation results validate the effectiveness of our feature extraction pipeline: the selected features carry hierarchically ordered membership privacy signals, and VLALeaks-MLP is more suitable for high-dimensional full-feature attack, while VLALeaks-RF is more robust for lightweight sparse-feature inference, providing further insight for lightweight practical attack deployment on VLA models.

\subsubsection{Ablation Experiments of Feature}

\begin{table*}[!t]
\centering
\caption{Attack AUC of VLALeaks under the defense of removing partial duplicate data. We test six tasks on $\pi_0$. Tasks (a)-(f) are described in Section~\ref{ddd}.}
\label{defense}
\begin{adjustbox}{width=0.65\textwidth}
\begin{tabular}{lcccccc}
\toprule
Method & (a) & (b) & (c) & (d) & (e) & (f)\\
\midrule
Loss & -0.0001 & -0.0001 & 0.0000 & 0.0000 & -0.0001 & 0.0000\\
Min-k & -0.0002 & 0.0000 & +0.0001 & 0.0000 & 0.0000 & +0.0001\\
Min-k++ & 0.0000 & -0.0000 & 0.0000 & 0.0000 & -0.0001 & 0.0000\\
MaxR-img & 0.0000 & 0.0000 & -0.0003 & 0.0000 & 0.0000 & 0.0000\\
MaxR-inst & 0.0000 & 0.0000 & 0.0000 & 0.0000 & 0.0000 & -0.0003\\
MaxR-desp & -0.0001 & 0.0000 & 0.0000 & 0.0000 & 0.0000 & 0.0000\\
SMI & 0.0000 & -0.0001 & 0.0000 & 0.0000 & 0.0000 & 0.0000\\
RIM & 0.0000 & 0.0000 & 0.0000 & 0.0000 & -0.0002 & 0.0000\\
RINM & 0.0000 & 0.0000 & 0.0000 & -0.0001 & 0.0000 & 0.0000\\
\midrule
VLALeaks-RF & 0.0000 & 0.0000 & 0.0000 & 0.0000 & -0.0001 & 0.0000\\
VLALeaks-MLP & 0.0000 & 0.0000 & +0.0001 & 0.0000 & 0.0000 & 0.0000\\
\bottomrule
\end{tabular}
\end{adjustbox}
\end{table*}

\begin{figure}[!t]
    \centering
    \begin{subfigure}[b]{0.76\linewidth}
        \centering
        \includegraphics[width=\linewidth]{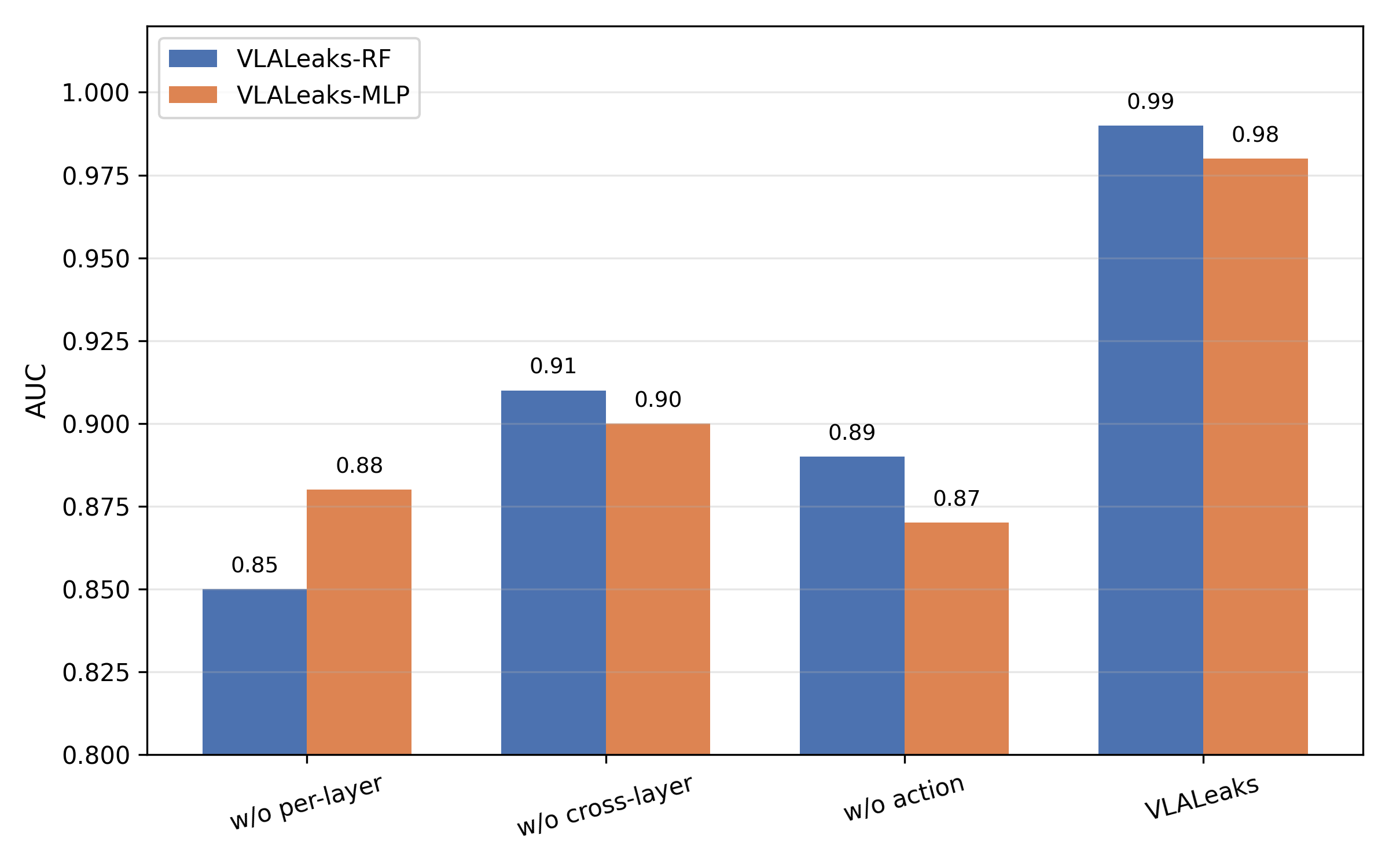}
        \caption{Pull the tray right.}
        \label{fig:ablation_study_a}
    \end{subfigure}
    \hfill
    \begin{subfigure}[b]{0.76\linewidth}
        \centering
        \includegraphics[width=\linewidth]{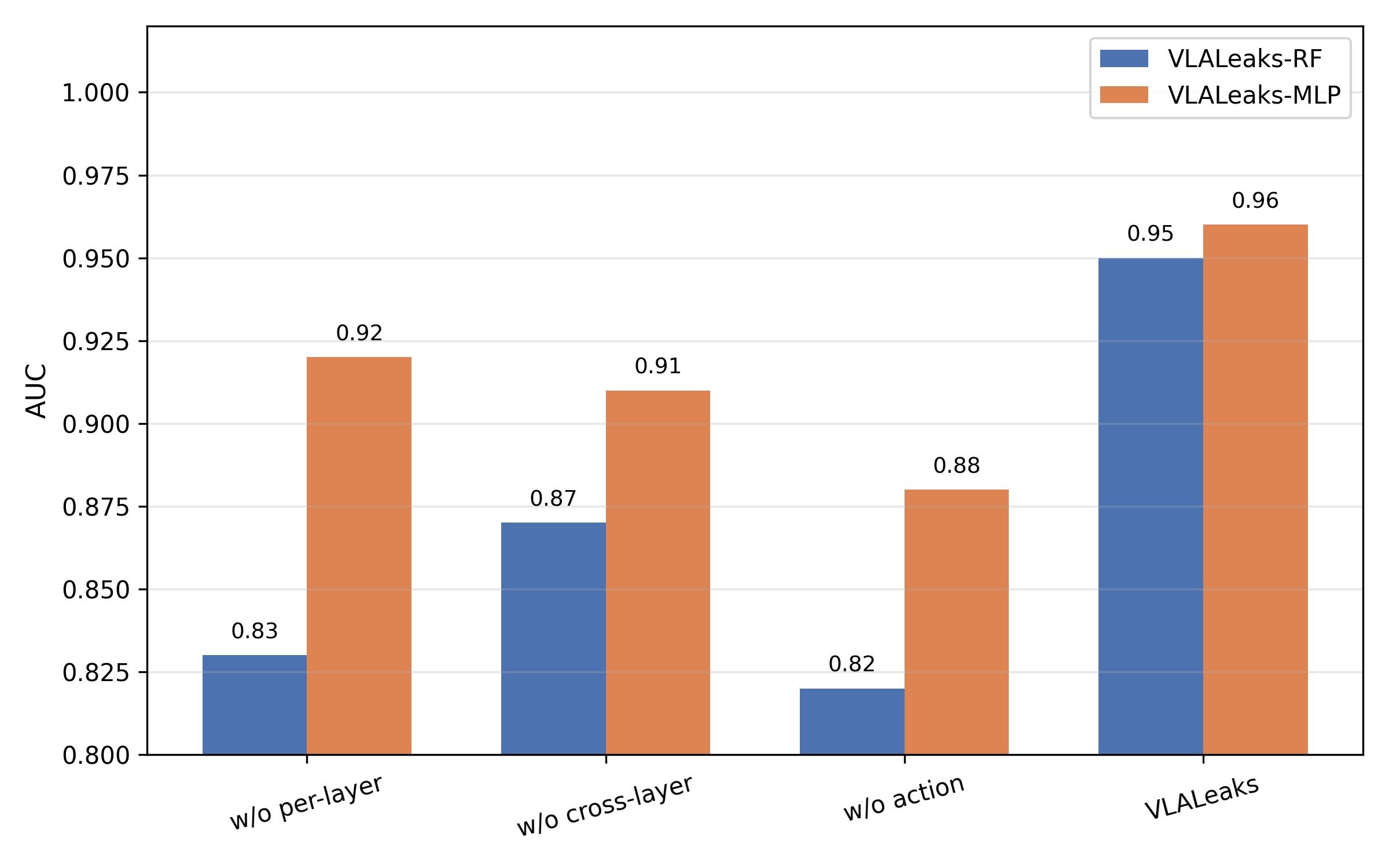}
        \caption{Pick up a candy and put it back.}
        \label{fig:ablation_study_d}
    \end{subfigure}
    \caption{Ablation study of attack AUC using three feature extraction groups. Experiments are performed on two tasks. Here, 'w/o' stands for 'without', indicating the removal of one feature group such that only the remaining two groups are used for MIAs.}
    \label{fig:ablation_combined}
\end{figure}

We perform ablation experiments in Figure~\ref{fig:ablation_combined} to quantify the individual contribution of three core feature branches (per-layer feature, cross-layer feature, action feature) in VLALeaks in two representative robotic manipulation tasks. Three ablated variants are set as: without per-layer, without cross-layer, and without action, where each setting discards one feature group and retains the other two for membership inference, and the full VLALeaks with all three feature types serves as the complete baseline.
Across both task (a) and task (d), removing any single feature component consistently leads to AUC degradation for both VLALeaks-RF and VLALeaks-MLP compared with the full-model baseline.
For task (a) Pull the tray right: Full VLALeaks achieves AUC=0.99 (RF)/0.98 (MLP). After feature ablation, performance drops to 0.85/0.88 (without per-layer), 0.91/0.90 (without cross-layer), 0.89/0.87 (without action). The most severe performance decline occurs when discarding per-layer features, demonstrating per-layer hidden features carry the richest membership privacy information for this task.
For task (d) Pick up a candy and put it back: The complete model reaches AUC=0.95 (RF)/0.96 (MLP). The AUC drops to 0.83/0.92 (without per-layer), 0.87/0.91 (without cross-layer), 0.82/0.88 (without action). Here, eliminating action features brings the largest accuracy loss, meaning that action-related features dominate privacy leakage on this candy-manipulation task.
Overall, all three feature modules deliver indispensable complementary privacy clues: none of the three feature subsets alone can replace the full feature combination. The task-dependent varying importance of each feature group also explains why aggregating per-layer, cross-layer, and action features simultaneously enables VLALeaks to capture comprehensive cross-modal privacy from heterogeneous VLA data and reach optimal attack performance.

\subsection{VLALeaks against Defenses}
We conduct defenses in Table~\ref{defense} to evaluate the resilience of VLALeaks against partial duplicate training data removal~\cite{kandpal2022deduplicating}, a prevalent dataset sanitization defense for mitigating model memorization risks. The table records the AUC change value (defense setting vs original baseline) of VLALeaks-RF and VLALeaks-MLP across six robot manipulation tasks.
As observed, all AUC fluctuations of two variants are restricted within the tiny range $\left [ -0.0001,+0.0001 \right ]$  across six tasks: VLALeaks-RF only obtains a negligible $-0.0001$ AUC drop on task (e) and remains unchanged for all remaining five tasks; VLALeaks-MLP merely has a marginal $+0.0001$ improvement on task (c) with zero variation on other cases. 
VLALeaks extracts implicit membership privacy from the inherent statistical and nonlinear distribution of vision-language-action latent features rather than recurring duplicated samples. Consequently, the duplicate-data removal defense fails to eliminate the privacy vulnerability exploited by our approach. Differential privacy~\cite{abadi2016deep} is widely adopted as a defense against MIAs. We attempt to incorporate it into VLA models. However, we find that differential privacy not only significantly degrades model performance but also fails to effectively defend against our proposed attack. The above results verify the outstanding anti-defense robustness of VLALeaks, demonstrating that the privacy leakage uncovered by our method is rooted in intrinsic VLA model characteristics.

\section{Conclusion}
\label{Conclusion}
In this work, we take the first step toward investigating membership privacy leakage in Vision-Language-Action (VLA) models, thereby addressing a critical gap in the literature. We formally define the membership inference attacks (MIAs) problem for VLA models and analyze the fundamental reasons why existing methods fail to achieve satisfactory performance. To address this, we propose VLALeaks, a novel MIA specifically tailored for VLA models. Through a carefully designed two-stage pipeline, VLALeaks achieves strong attack AUC and TPR@1\%FPR, approaching near-perfect inference under certain configurations. Moreover, our two-stage attack reveals the memorization of member information within attention matrices, providing new insights into privacy protection for VLA models.

We conduct extensive and rigorous experiments in both simulated and real-world robotic environments, demonstrating the effectiveness of VLALeaks. Our attack exposes previously unidentified privacy leakage blind spots in current VLA models, underscoring their vulnerability. We hope that our work will inspire future research on privacy-preserving VLA models and inform the development of privacy auditing tools for robotics systems, paving the way toward secure, trustworthy VLA models.

\newpage






\bibliographystyle{IEEEtran} 
\bibliography{ref}
%



\section*{Ethics Considerations}
This paper investigates membership inference attacks on VLA models. All experiments were conducted using publicly available datasets and standard open-source model architectures, none of which involve direct ethical concerns. Our work reveals the substantial and previously underestimated privacy risks posed by VLALeaks, underscoring the critical importance of this emerging threat vector. We aim to stimulate both academia and industry to develop more robust defense mechanisms against these evolving threats.

\appendix

\subsection{Additional Experiment Results}
\label{AdditionalExperiment}

\begin{figure}[!t]
    \centering
    \includegraphics[width=0.52\linewidth]{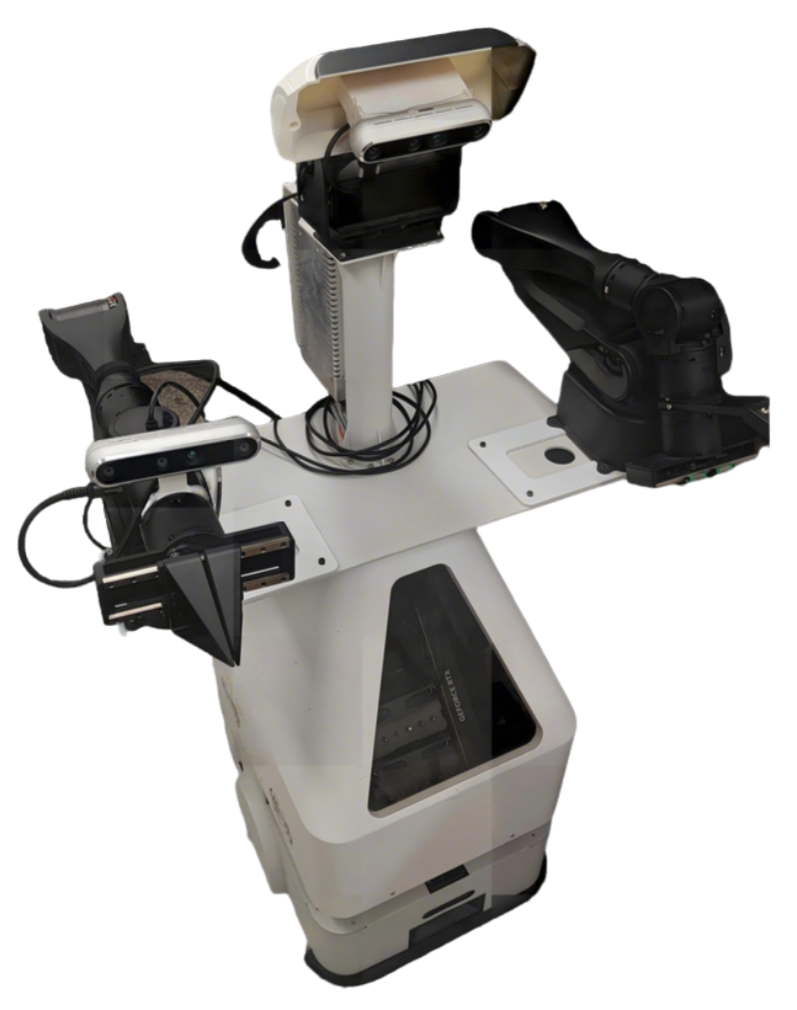}
    \caption{Real-world mobile manipulation platform. The system comprises an AIRBOT Play arm mounted on a Woosh wheeled base, with head- and wrist-mounted RGB-D cameras providing visual observations. All onboard computation is housed inside the base.}
    \label{fig:platform2}
\end{figure}

To further validate the effectiveness of VLALeaks, we fine-tune VLA models on downstream real-world data and then perform MIAs. Specifically, we deploy $\pi_0$~\cite{black2024pi_0} on a mobile manipulation platform consisting of a single AIRBOT Play arm coupled with a Woosh wheeled base (Figure~\ref{fig:platform2}). This platform enables tasks that require the robot to navigate between locations and perform contact-rich manipulation upon arrival.

\subsubsection{Task Definition and Data Collection}

\begin{figure}[!t]
    \centering
    \includegraphics[width=0.66\linewidth]{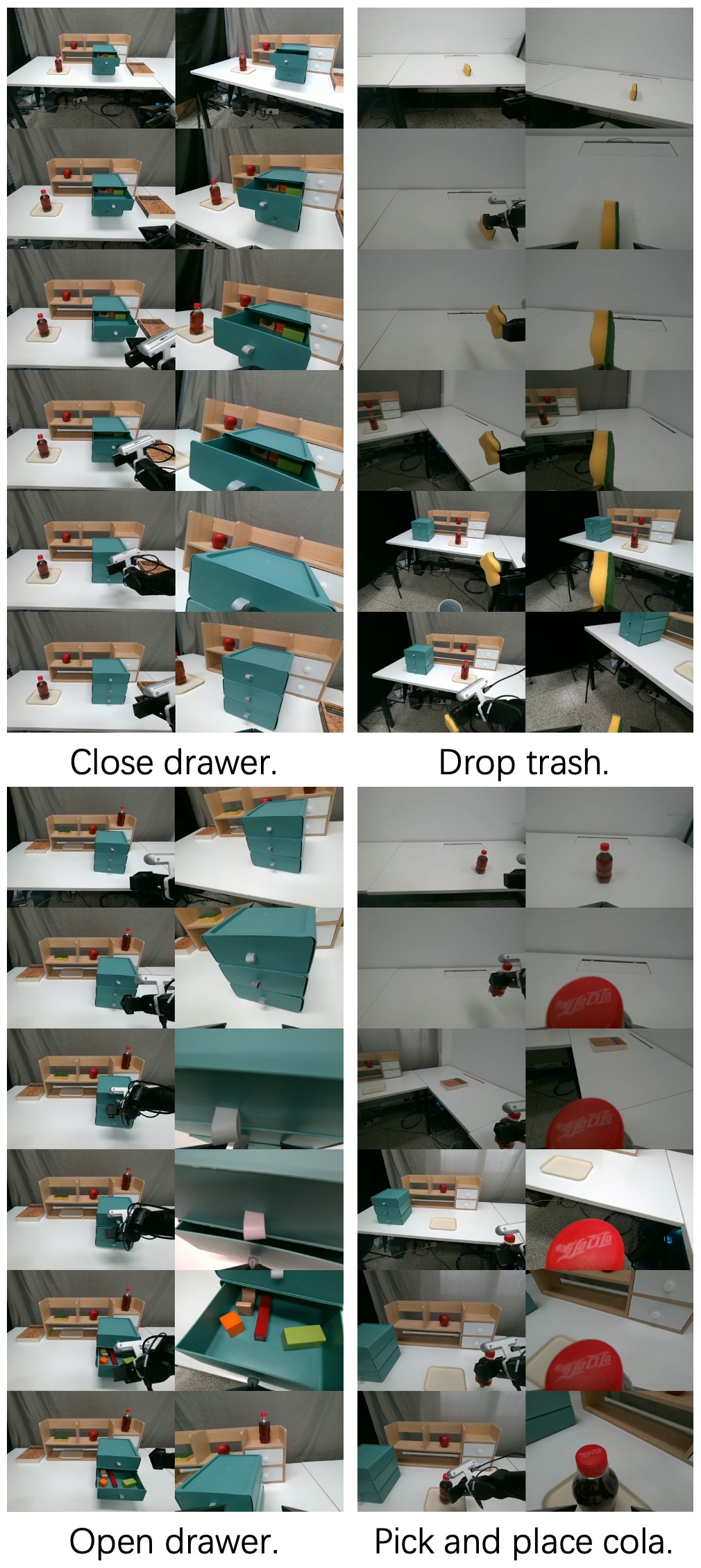}
    \caption{Four tasks for the mobile manipulation platform with the head camera (left) and the wrist camera (right).}
    \label{fig:robot2}
\end{figure}

As illustrated in Figure~\ref{fig:robot2}, we design four tasks encompassing both articulated-object interaction and long-horizon pick-transport-place behaviors. Each task requires tight coordination between base locomotion and arm control, as opposed to manipulation from a fixed stance:
\begin{itemize}
    \item (aa) \textbf{Close drawer.} Each episode starts with the robot positioned away from the workstation. The base drives forward until the robot reaches the front of the table; then the arm extends to push the drawer shut. A trial is considered successful if the drawer is fully closed upon arm retraction.
    \item (bb) \textbf{Drop trash.} The base advances to the table edge, and the arm reaches out to grasp a target sponge block. The robot then rotates leftward and drives toward the trash bin. Once at an appropriate standoff distance, the arm carries the sponge above the bin opening and releases the gripper to dispose of it.
    \item (cc) \textbf{Open drawer.} The robot starts from a remote position, navigates to the table front, and then clamps the arm onto the drawer handle to slowly pull it outward. Success is achieved if the drawer opens without the gripper slipping off the handle.
    \item (dd) \textbf{Pick and place the cola.} The base moves to the table, and the arm grasps a bottle of cola. The robot then turns left and travels to a second table equipped with a receiving tray. Once in a suitable pose, the arm lowers the bottle onto the tray and releases it.
\end{itemize}

All demonstrations are gathered through whole-body teleoperation, with the operator jointly commanding base motion and arm trajectories so that the recorded data reflects naturally coupled locomotion–manipulation behavior. Demonstrations are logged at the same 10\,Hz frequency used during autonomous execution to ensure temporal consistency between training and deployment.


\subsubsection{Mobile Manipulation Platform}

The platform integrates a 6-DOF AIRBOT Play arm onto a Woosh omnidirectional wheeled base. The arm operates in joint position control mode and is equipped with a parallel-jaw 1-DOF gripper (stroke: 0–60\,mm). The Woosh base accepts linear and angular velocity commands. To maintain mechanical stability and safe operation when the arm is extended, we limit the base to a maximum linear speed of 0.2\,m/s and a maximum angular speed of 0.3\,rad/s during both data collection and policy rollout.


For teleoperated data collection, an AIRBOT Replay arm serves as the leader device driving the AIRBOT Play follower arm, while base velocities are commanded concurrently through the same interface. The leader–follower arm pair is identical to that used in our fixed-base setup, enabling demonstration pipelines to be reused across platforms. Visual observations come from two RGB-D cameras mounted on the platform: one at the head, providing a broad forward-facing view for base navigation and coarse target localization, and one at the wrist, offering a close-up view of the end-effector workspace for fine manipulation. This head-plus-wrist arrangement supplies complementary global and local visual context throughout each task.




\subsubsection{Policy Training}

We train task-specific policies on the mobile platform using $\pi_0$~\cite{black2024pi_0}. The policy consumes synchronized RGB streams from the head and wrist cameras together with the robot's proprioceptive state. At each control step, it predicts a 9-dimensional whole-body action vector composed of six arm joint targets, one gripper command, and two base velocity terms (linear and angular). Because the deployed platform exposes neither torso-lift nor head articulation, the action space omits these degrees of freedom, yielding a compact command interface tailored to this hardware. Policies are executed in a receding-horizon manner at 10\,Hz, matching the control rate used during demonstration logging.

\begin{table}[!t]
    \centering
    \caption{Attack AUC on $\pi_0$ across four real-world mobile manipulation tasks (aa-dd).}
    \begin{tabular}{lcccc}
        \toprule
        \textbf{Method} & \textbf{(aa)} & \textbf{(bb)} & \textbf{(cc)} & \textbf{(dd)} \\
        \midrule
        Loss & 0.52 & 0.47 & 0.50 & 0.49 \\
        VLALeaks-RF & 0.89 & 0.83 & 0.85 & 0.90 \\
        VLALeaks-MLP & 0.87 & 0.81 & 0.82 & 0.94 \\
        \bottomrule
    \end{tabular}
    \label{tab:comparison1}
\end{table}

\subsubsection{Attack AUC}
Table~\ref{tab:comparison1} presents the attack AUC for our proposed VLALeaks and Loss attack~\cite{yeom2018privacy} across four real-world mobile manipulation tasks to validate its real-world effectiveness. Loss attack is close to random guessing. Both VLALeaks-RF and VLALeaks-MLP achieve robust and high AUC values above 0.81 across all physical robot tasks, which solidly demonstrates that both instantiations of our VLALeaks are effective at uncovering membership privacy in real-deployed VLA models. For the tasks (aa, bb, cc), VLALeaks-RF achieves strong performance with AUC ranging from 0.83 to 0.89, while VLALeaks-MLP maintains competitive AUC between 0.81 and 0.87. On the task (dd), both methods attain their peak performance: VLALeaks-RF reaches 0.90, and VLALeaks-MLP obtains an even higher AUC of 0.94. Regardless of task type and complexity, neither VLALeaks suffers dramatic performance drops, confirming that both attack classifiers successfully exploit the multimodal privacy features extracted by our VLALeaks. These real-world results further corroborate the reliability and practical efficacy of our VLALeaks.
\end{document}